\documentclass[11pt,a4paper]{article}
\usepackage{jheppub}

\usepackage{lmodern}
\usepackage[utf8]{inputenc}
\usepackage{color}
\usepackage{float}
\usepackage{amsmath,bm}
\usepackage{amssymb}
\usepackage{graphicx}
\usepackage{setspace}
\usepackage{subfigure}
\usepackage{amsthm}
\usepackage{amsfonts}
\usepackage{braket}
\usepackage{multirow}
\usepackage{ulem}
\usepackage{slashed}
\usepackage{babel}
\usepackage{lipsum}
\usepackage{hyperref}
\usepackage{url}
\hypersetup{
	colorlinks=true,
	linkcolor=cyan,
	urlcolor=blue,
	citecolor=red,
}

\title{Entanglement of defect subregions in double holography}

\author[a,b]{Yuxuan Liu,}

\author[c]{Qian Chen,}

\author[c,d]{Yi Ling,}

\author[b]{Cheng Peng,}

\author[c,e]{Yu Tian,}

\author[f]{and Zhuo-Yu Xian}

\affiliation[a]{
Institute of Quantum Physics, School of Physics, Central South University, Changsha 418003, China
}

\affiliation[b]{Kavli Institute for Theoretical Sciences (KITS), University of Chinese Academy of Sciences, Beijing 100190, China}

\affiliation[c]{School of Physical Sciences, University of Chinese Academy of Sciences, Beijing 100049, China}

\affiliation[d]{Institute of High Energy Physics, Chinese Academy of Sciences, Beijing 100049, China}

\affiliation[e]{Institute of Theoretical Physics, Chinese Academy of Sciences,
Beijing 100190, China}

\affiliation[f]{
Institute for Theoretical Physics and Astrophysics and W\"urzburg-Dresden Cluster of Excellence ct.qmat, Julius-Maximilians-Universit\"at W\"urzburg, 97074 W\"urzburg, Germany}

\emailAdd{alanfang@csu.edu.cn}
\emailAdd{chenqian192@mails.ucas.ac.cn}
\emailAdd{lingy@ihep.ac.cn}
\emailAdd{pengcheng@ucas.ac.cn}
\emailAdd{ytian@ucas.ac.cn}
\emailAdd{zhuo-yu.xian@physik.uni-wuerzburg.de}

\abstract{
In the framework of double holography, we investigate the entanglement behavior of a brane subregion in AdS spacetime coupled to a bath on its boundary and also extract the contribution from the quantum matter within this subregion. From the boundary perspective, the brane subregion serves as the entanglement wedge of a subsystem on the conformal defect. In the ground state, we find the subsystem undergoes an entanglement phase transition induced by the degrees of freedom on the brane. With subcritical parameters, the wedge and entanglement entropy sharply decrease to zero. In contrast, in the supercritical regime, both the wedge and entropy stabilize, enabling analysis of both entanglement and reflected entropy. In this phase, we derive formulas for entanglement measures based on defect and bath central charges in the semi-classical limit. For entanglement entropy, the classical geometry only contributes a subleading term with logarithmic divergence, but the brane-bath CFT entanglement exhibits a dominant linear divergence, even in the semi-classical limit. Regarding reflected entropy within the defect subsystem, classical geometry contributes a leading term with logarithmic divergence, while the quantum matter within the entanglement wedge only contributes a finite term.
}

\begin{document}

\maketitle

\section{Introduction}
Entanglement entropy, pivotal in quantum many-body systems and quantum information theory, quantifies quantum entanglement in bipartite pure states. Its extensive study is crucial for uncovering the physics underpinning these systems.

The entanglement of a subsystem $\mathcal{A}$ with its environment can be {measured by} von Neumann entropy $S_\mathcal{A} = -\text{Tr}(\rho_\mathcal{A} \log \rho_\mathcal{A})$, where $\rho_\mathcal{A}$ is the reduced density matrix of $\mathcal{A}$. In the context of {AdS/CFT}  correspondence \cite{Maldacena:1997re,Gubser:1998bc,Witten:1998qj}, the classical spacetime in the bulk corresponds to a quantum state of the boundary system. The Ryu-Takayanagi (RT) formula \cite{Ryu:2006bv,Ryu:2006ef}, along with its covariant version \cite{Hubeny:2007xt}, links the entanglement entropy of a boundary subsystem $\mathcal{A}$ to the area of a minimal (or extremal) surface $\gamma_\mathcal{A}$ in the bulk, which is homologous to the subregion $\mathcal{A}$ and shares the same boundary with it.
Moreover, {once the degrees of freedom of quantum fields in the bulk are} considered, the RT surface should be generalized to a quantum extremal surface (QES), represented as $\gamma_{\mathcal{A}}$. The holographic entanglement entropy (HEE) of $\mathcal{A}$ is calculated by extremizing the generalized gravitational entropy functional \cite{Lewkowycz:2013nqa,Engelhardt:2014gca,Witten:2021unn},
\begin{equation}\label{eq:QESinRad}
S[\mathcal{A}]=\min \left\{ \text{ext} \left[\frac{\operatorname{Area}[\gamma_\mathcal{A}]}{4 G^{d}_{N}} + S_{\text{QFT}}\right]\right\},
\end{equation}
where $G^{d}_{N}$ denotes the Newton constant in the $d$-dimensional asymptotically Anti-de Sitter (AAdS) spacetime, and $S_{\text{QFT}}$ represents the entanglement entropy of the quantum fields enclosed by $\gamma_\mathcal{A}$ and the boundary region $\mathcal{A}$. 
A key application of the QES formula arises in describing the gravity-plus-bath system. Here, a $d$-dimensional AAdS gravity theory is coupled to a $d$-dimensional flat bath on its boundary \cite{Almheiri:2019hni,Almheiri:2019psf,Almheiri:2019yqk,Penington:2019kki,Almheiri:2019qdq,Almheiri:2019psy}. Quantum fields in both spacetimes, modeled by the same conformal field theory (CFT), have their entanglement entropy computed via the island formula \cite{Almheiri:2019hni,Chen:2019uhq},
\begin{equation}\label{eq:QESinRad2}
S[\mathcal{R}]=\min_{\mathcal I} \left\{ \mathop{\text{ext}}\limits_{\mathcal I} \left[\frac{\operatorname{Area}[\partial \mathcal{I}]}{4 G_{N}} + S[\mathcal{R} \cup \mathcal{I}]\right]\right\}.
\end{equation}
Here $\mathcal{R}$ and $\mathcal{I}$ denote the quantum fields in the bath and the islands, respectively. This perspective is commonly referred to as the brane perspective.

Computing the entanglement entropy of quantum fields often presents challenges. Yet, in cases where these fields align with a CFT of large central charge, holographic duality comes into play. This duality posits that a $d$-dimensional bulk corresponds to a Planck brane in higher dimensions, allowing for a geometric interpretation of the entropy formula (\ref{eq:QESinRad2}). This can be further detailed as \cite{Chen:2020uac,Chen:2020hmv,Hernandez:2020nem,Grimaldi:2022suv}, 
\begin{align}\label{eq:Intro_QES}
S[\mathcal R]= \min_{\mathcal I} \left\{\mathop{\text{ext}}\limits_{\mathcal I} \left[\frac{1}{4G_N^{(d+1)}}\textbf{Area}(\gamma_{\mathcal I\cup \mathcal{R}}) + \frac{1}{4G_b^{(d)}} \,\textbf{Area}(\partial\mathcal I)\right]\right\},
\end{align}
where the first term arises from the area of the $(d-1)$-dimensional RT surface bounded by $\mathcal{I} \cup \mathcal{R}$, and $G_b$ is the Newton constant introduced by the DGP term.\footnote{The island formulae (\ref{eq:QESinRad2}) and (\ref{eq:Intro_QES}) are pivotal in addressing the black hole information paradox within holography \cite{hawking1974black,hawking1975particle,hawking1976breakdown,Page:1993wv,susskind1993stretched,Page:2004xp,Almheiri:2012rt,Page:2013dx,Hartman:2013qma}. On {one} hand, the concept of higher-dimensional islands was introduced in \cite{Almheiri:2019psy}, replacing the lower-dimensional one black hole with a tensed brane under Neumann boundary conditions \cite{Takayanagi:2011zk,Chu:2018ntx,Miao:2018qkc}. This approach requires sufficient degrees of freedom (DOF) on the brane to derive a dynamical Page curve \cite{Ling:2020laa,Geng:2020fxl,Geng:2020qvw}, with various methods proposed for incorporating these DOF. Further explorations in higher-dimensional brane-world constructions are detailed in \cite{Krishnan:2020fer,Miao:2020oey,Akal:2020wfl,Akal:2020twv,Omidi:2021opl}.

On the other hand, path integral formalism reveals that islands emerge due to another saddle point's dominance, {namely} the wormhole saddle point, where Euclidean wormholes connect different replicas \cite{Penington:2019kki,Almheiri:2019qdq} (also see \cite{Rozali:2019day,Karlsson:2020uga}). This viewpoint spurred extensive discussions on wormholes as bridges {among} disjoint universes \cite{Balasubramanian:2020xqf,Balasubramanian:2021wgd,Balasubramanian:2020coy,Miyata:2021ncm,Miyata:2021qsm}, lengthening behind horizons, and {on}  baby universes connecting to parent universes through wormholes \cite{Marolf:2020rpm,Marolf:2020xie,Balasubramanian:2020jhl,Peng:2021vhs,Peng:2022pfa}. Recent discussions in this field are found in \cite{Alishahiha:2020qza,Hashimoto:2020cas,Anegawa:2020ezn,Hartman:2020swn,Chen:2020jvn,Bhattacharya:2020uun,Deng:2020ent,Wang:2021woy,He:2021mst,Gautason:2020tmk,Krishnan:2020oun,Sybesma:2020fxg,Chou:2021boq,Hollowood:2021lsw,Suzuki:2022xwv,Suzuki:2022yru,Bhattacharya:2021nqj,Bhattacharya:2021dnd,Caceres:2021fuw,Bhattacharya:2021jrn,Caceres:2020jcn,Chen:2019iro,Balasubramanian:2020hfs,Almheiri:2020cfm,Li:2020ceg,KumarBasak:2020ams,Anderson:2020vwi,Vardhan:2021mdy,Kawabata:2021vyo,Kawabata:2021hac,Geng:2021iyq,Geng:2021mic,Akal:2021dqt,Renner:2021qbe,Balasubramanian:2021xcm,Engelhardt:2022qts,Afrasiar:2022ebi}.}

{This formula indicates that the entanglement entropy is obtained} 
by extremizing over all possible islands, {and then choosing that} one with {the} minimal area. This approach forms the bulk perspective, paralleling the brane perspective from alternate angles \cite{Almheiri:2019hni}. Furthermore, the gravity-plus-bath system corresponds to a $d$-dimensional boundary CFT (BCFT${}_d$) hosting a $(d-1)$-dimensional CFT living on its boundary. This CFT$_{d-1}$ may be regarded as a conformal defect \cite{Chen:2020uac}.

Nevertheless, {the investigation on the entanglement} over finite subregions is limited in existing literature, largely due to the complexity of constructing extremal surfaces near the Planck brane. {In} 
the ground state for $4$-dimensional gravity-plus-bath theory, the formation of entangling surfaces with various shapes $\mathcal{A}$ on the conformal boundary has been facilitated by the ``Surface Evolver'' tool \cite{Fonda:2014cca,Fonda:2015nma,Seminara:2017hhh,Seminara:2018pmr,Cavini:2019wyb}.
Extensively employed in {the field of material science}, this tool facilitates the evolution of a surface to its minimal energy state via gradient descent methods \cite{brakke1992surface,evolverlink}.
In scenarios where a single subregion $\mathcal{A}$ intersects with the boundary of the Planck brane -- Fig.~\ref{fig:regionA}, the leading divergent term in the entanglement entropy of $\mathcal{A}$ aligns proportionally with the boundary area of $\mathcal{A}$, reflecting the area law. Conversely, the secondary divergent term adheres to a logarithmic law:
\begin{equation}
    S[\mathcal{A}]=\frac{\textbf{Area}(\partial\mathcal{A})}{\epsilon}+\mathcal{G} \log \frac{\textbf{Area}(\partial\mathcal{A})}{\epsilon}+\mathcal{O}(1),
\end{equation}
where $\epsilon$ represents the UV cut-off in the $d$-dimensional CFT. The coefficient $\mathcal{G}$, known as the corner function, links with the angle at which $\partial\mathcal{A}$ intersects the brane. This corner function illuminates the behavior of the entanglement entropy near the boundary of the brane, offering insights into the geometric attributes of the entangling surface. Despite its complexity, the calculation of the subleading divergent term can be performed using the Willmore functional with an appropriate boundary term on the extremal surface \cite{Seminara:2018pmr}.


Within the doubly holographic framework, a fascinating area of study is the role of quantum matter in entanglement within semi-classical gravity. Especially from the brane perspective, we can investigate the entanglement properties of the brane subregion by calculating the QES. From the boundary perspective, the brane subregion can serve as the (quantum) entanglement wedge of a subsystem on the conformal defect. To gradually lessen the influence of the bath CFT, we design a series of specific subsystems. By reducing their areas while maintaining the size of the defect subsystem, we can finally extract the entanglement from the quantum matter within a finite brane subregion. In specific, we will calculate the $(d-1)$-dimensional RT surface from the perspective of bulk gravity. Due to the {presence of} Planck brane and the finite subregion $\mathcal{A}$, this RT surface lacks translational invariance along two spatial directions -- Fig.~\ref{fig:QES}. However, when constructing the RT surface of $\mathcal{D}$, the numeric solution from ``surface evolver'' is usually unstable \cite{Seminara:2017hhh}. Therefore, we directly solve second-order partial differential equations, adhering to Neumann boundary conditions on the Planck brane and Dirichlet conditions on the conformal boundary. We employ pseudo-spectral and finite difference methods combined with Newton-Raphson iterations for the numerical resolution of these equations.

We organize this paper as follows: In Section~\ref{sec:2}, we introduce the doubly holographic setup and focus on the RT surface against the vacuum background. 
In Section~\ref{sec:3}, we numerically analyze the holographic entanglement entropy 
 and reflected entropy of some explicit subregions and distinguish contributions from gravity and quantum fields. 
Finally, in Section~\ref{sec:4}, we conclude the paper and discuss the future directions.

\section{The setup in double holography} \label{sec:2}
We consider a $(d+1)$-dimensional AAdS spacetime intersected by a $d$-dimensional Planck brane $\mathcal{B}$. The brane intersects with the conformal boundary $\bm{\partial}$ at a $(d-1)$-dimensional intersection denoted by $\partial\mathcal{B}$ \cite{Almheiri:2019hni,Chen:2020uac,Almheiri:2019yqk,Takayanagi:2011zk}. There exist three equivalent descriptions of this system from different perspectives.

\subsection{The gravitational action}
In the context of the double holography \cite{Almheiri:2019hni,Chen:2020uac}, the action of the $(d+1)$-dimensional gravity theory can be expressed as:

\begin{align}\label{eq:Action}
I=
&\frac{1}{16\pi G_N^{(d+1)}} \Bigg[ \int d^{d+1}x
\sqrt{-g}\left(R+\frac{d(d-1)}{L^2}\right)+2\int_{\bm{\partial}}d^{d}x\sqrt{-h_{\bm{\partial}}}K_{\bm{\partial}}\nonumber\\
& + 2 \left(\int_{\mathcal{B}}d^{d}x\sqrt{-h}\left(K-\alpha\right)-\int_{\bm{\partial}\mathcal{B}}
d^{d-1}x \sqrt{-\Sigma}\, (\pi-\theta_0)\right)\Bigg]\nonumber \\
&+ \left[\frac{1}{16 \pi G_{b}^{(d)}}\int_{\mathcal{B}} d^d x
\sqrt{-h}R_{h}+\frac{1}{8\pi G_{b}^{(d)}}\int_{\mathcal{B}\cap \bm{\partial}} d^{d-1}x \sqrt{-\Sigma} \, k\right].
\end{align}
In this expression, $G_N^{(d+1)}$ and $L$ denote the Newton constant and AdS radius in the bulk, respectively. $K_{\bm{\partial}}$ and $h_{\bm{\partial}}$ are the extrinsic curvature and induced metric on the conformal boundary $\bm{\partial}$, while $K$, $h$, and $\alpha$ represent the extrinsic curvature, induced metric, and tension term on the Planck brane $\mathcal{B}$. The induced metric and inner angle between $\mathcal{B}$ and $\bm{\partial}$ on the intersection $\partial\mathcal{B}$ are denoted by $\Sigma$ and $\theta_0$, respectively. Additionally, a Dvali-Gabadadze-Porrati (DGP) term \cite{Randall:1999vf,Dvali:2000hr} is introduced in the last line, with $G_b^{(d)}$ representing the Newton constant on the brane, $R_h$ being the intrinsic curvature on $\mathcal{B}$, and $k$ denoting the additional extrinsic curvature at the intersection.

This perspective is commonly referred to as the bulk perspective, and there exist two equivalent descriptions of the aforementioned gravity system:
\begin{description}
    \item [Brane perspective] The $(d+1)$-dimensional gravity in the bulk is dual to a semi-classical gravity on the brane $\mathcal{B}$, with CFT living on both the brane $\mathcal{B}$ and the heat bath $\bm{\partial}$ \cite{Gubser:1999vj}.
    \item [Boundary perspective]  The gravity-plus-bath theory on the brane can further be dual to the boundary conformal field theory (BCFT) coupled to a conformal defect on its boundary \cite{Takayanagi:2011zk,Almheiri:2019hni,Chen:2020uac}.
\end{description}

The main construction presented in this paper is elaborated on from the bulk gravity perspective. In order to realize dynamical gravity on the brane, we impose Neumann boundary conditions on the Planck brane (see also \cite{Miao:2017gyt,Chu:2017aab,Miao:2018qkc} for alternative viewpoints), which are given by:
\begin{equation}\label{eq:BCSonBrane_General}
  K_{ij}-K h_{ij}+\alpha h_{ij}=\lambda L \left[\frac{1}{2}R_h h_{ij}-(R_{h}){}_{ij}\right], \quad \lambda=G_{N}^{(d+1)}/(G_{b}^{(d)}L),
\end{equation}
where $h_{ij}$ is the metric on the brane $\mathcal{B}$.

In this work, we focus on the simplest example, the construction on pure $\text{AdS}_{4}$ spacetime containing a Planck brane, which is dual to the ground state. The corresponding metric in Poincare coordinate system $(z,x,y,t)$ is given by
\begin{align}\label{eq:AdSmetric}
ds^2=L^2\frac{-dt^2+dz^2+dx^2+d y^2}{z^2},
\end{align}
where $L$ is the AdS radius. We can use a set of foliation coordinates, for numerical convenience, given by
\begin{equation}\label{eq:r_theta_y}
    z=r\sin\theta,\quad x=r\cos\theta,
\end{equation}
within the range
$$0 \leq r,\quad 0\leq\theta\leq\theta_0.$$
Note $\theta$ parametrizes the foliation of the AdS${}_4$ spacetime, which can be seen by substituting $x=-z\cot\theta$ into (\ref{eq:AdSmetric}). 
According to these coordinates, the spacetime metric can be expressed as
\begin{align}\label{eq:metric_in_theta_y}
ds^2=\frac{L^2 \csc^2\theta}{r^2}\left(-dt^2+dr^2+r^2d\theta^2 +dy^2\right),
\end{align}
with the conformal boundary $\bm{\partial}$ and the brane $\mathcal{B}$ being located respectively at 
\begin{align}
\partial: &\quad \theta=0,\\
\mathcal{B}:&\quad \theta=\theta_0.
\end{align}
The Neumann boundary condition for identifying the tension on the brane (\ref{eq:BCSonBrane_General}) can be written as\footnote{Different boundary conditions have been proposed in \cite{Seminara:2017hhh}.}
\begin{equation}\label{eq:BCSonBrane_AdS}
    \alpha L +\lambda \sin^2 \theta_{0}+2\cos \theta_0 = 0.
\end{equation}

With these constructions, the bulk geometry can be fully determined. Next, we will elaborate on the construction of RT surfaces in subregions, which capture the entanglement properties of finite-sized systems.
\subsection{Constructions on the RT surface}

\begin{figure}
  \centering
  \subfigure[]{\label{fig:bcft}
  \includegraphics[height=0.33\linewidth]{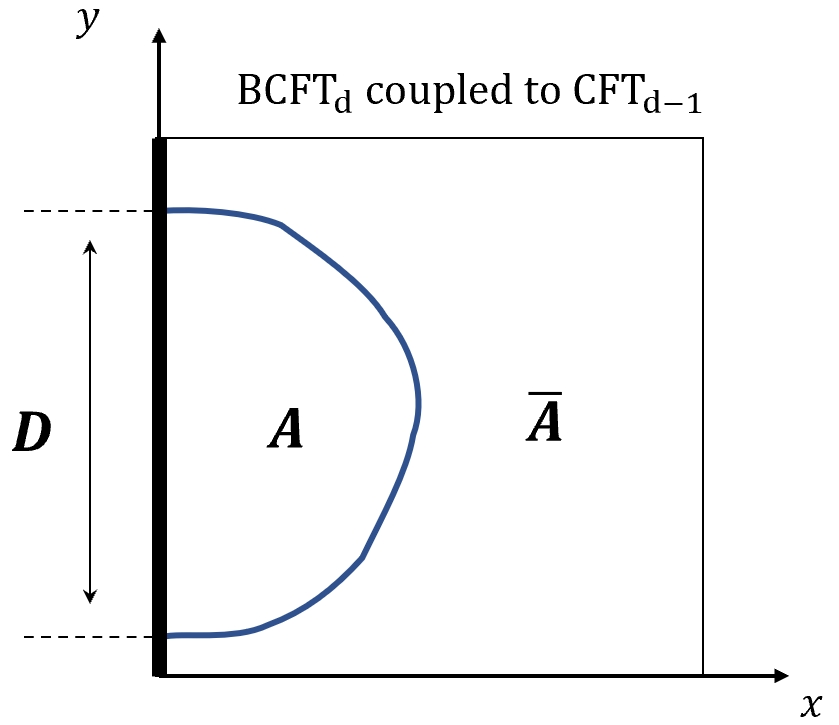}}
    \hspace{0pt}
  \subfigure[]{\label{fig:adscft}
  \includegraphics[height=0.33\linewidth]{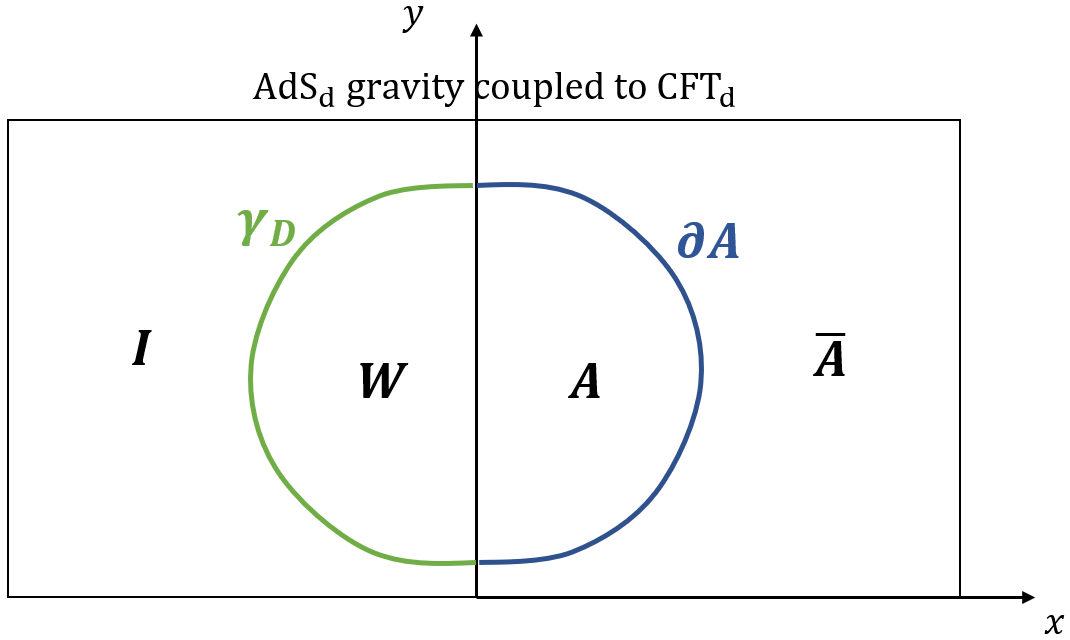}}
\caption{(a) The bipartition of a subsystem $\mathcal{A}$ (together with part of the conformal defect on its boundary, labeled as $\mathcal{D}$) and its complement. (b) The corresponding QES $\gamma_\mathcal{D}$ of a subsystem $\mathcal{W}\cup\mathcal{A}$ in the gravity-plus-bath theory, where $\mathcal{W}$ is the entanglement wedge of subsystem $\mathcal{D}$.}\label{fig:regionA}
\end{figure}
The entanglement entropy can also be viewed from the aforementioned three perspectives. First, from the boundary perspective -- as shown in Fig.~\ref{fig:bcft} --  we consider a subsystem $\mathcal{A}$, where the entire system is described by a specific state. The entanglement entropy of this subsystem is contributed both from $\mathcal{A}$ and part of the defect $\mathcal{D}=\mathcal{A}\cap\bm{\partial}\mathcal{B}$ on its boundary. Second, from the brane perspective -- Fig.~\ref{fig:adscft}, the corresponding subsystem becomes the union of $\mathcal{A}$ from the bath and $\mathcal{W}$ from the brane, which shares the same boundary with the quantum extremal island $\mathcal{I}$. Here the size of $\mathcal{W}$ is fully determined by the extremization process of the island, and the entanglement entropy is captured by the QES. Third, from the bulk perspective, the QES can be further dual to a standard HRT surface in the higher-dimensional bulk spacetime. The entanglement entropy can be described by the island formula, which can be expressed as follows:
\begin{align}\label{eq:QES}
S[\mathcal{A}]=\frac{L^{d-1}}{4G_N^{(d+1)}} \min_{\gamma_\mathcal{D}} \left\{\mathop{\text{ext}}\limits_{\gamma_\mathcal{D}} \left[\textbf{Area}(\Gamma_{\mathcal{W}\cup \mathcal{A}}) + \lambda L \,\textbf{Area}(\gamma_\mathcal{D})\right]\right\}.
\end{align}

The expression can be calculated after extremizing all possible $\gamma_\mathcal{D}$ and selecting the one with the minimal area. The first term represents the area of the classical extremal surface $\Gamma_{\mathcal{W}\cup \mathcal{A}}$ that penetrates into the bulk, while the last term is contributed by the DGP term.
Explicitly, the expression of the reduced entropy formula is defined to be
\begin{align}\label{eq:QES2}
S[\mathcal{A}]=\int d\sigma \mathcal{L}_{\text{bulk}} +\lambda L \int_{\mathcal{B}} d\sigma' \mathcal{L}_{\text{DGP}}
   ,
\end{align}
where $\frac{L^{d-1}}{4G_N^{(d+1)}}$ has been set to one for numerical convenience, $\sigma$ and $\sigma'$ denote the general coordinates on $\Gamma_{\mathcal{W}\cup \mathcal{A}}$ and $\gamma_\mathcal{D}$, respectively. 

To solve the entire system, we need to specify the corresponding boundary conditions. Generally, the shape of the subsystem $\mathcal{A}$ on the conformal boundary is predetermined, depending on the system of interest. Meanwhile, the entanglement wedge $\mathcal{W}$ of $\mathcal{D}$ can only be determined after the extremization process as mentioned in (\ref{eq:QES}). Therefore, we need to impose Neumann boundary conditions at $\gamma_\mathcal{D}$, whose locus is determined by the aforementioned extremization process. 
With this construction, we can calculate the entanglement entropy of a general subregion $\mathcal{W}\cup\mathcal{A}$ on gravitational backgrounds that support branes. Then, we will apply this construction to explicit examples.

\begin{figure}[h]
  \centering
    \includegraphics[width=250pt]{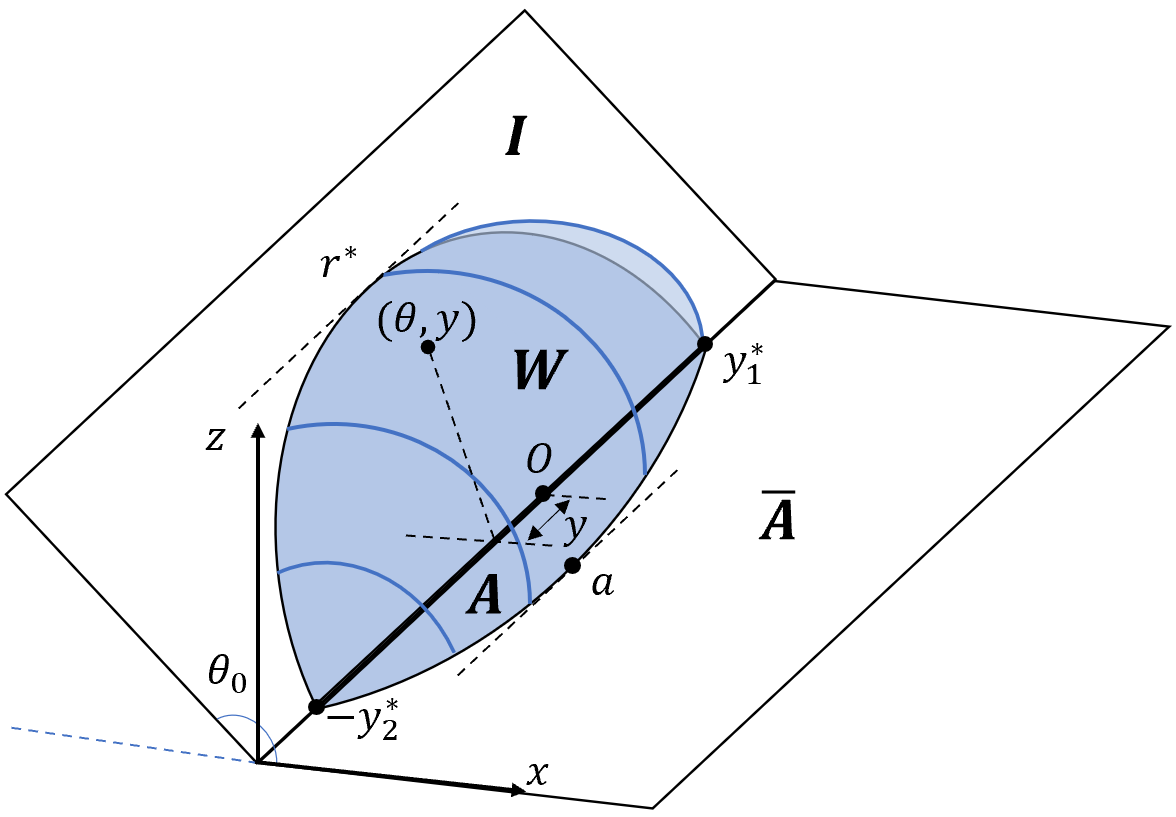}
  \caption{The holographic dual of QES for a finite subregion. The entanglement wedge of the subsystem $\mathcal{W}\cup\mathcal{A}$ is enclosed in the shaded region. Note that $a$ and $r^*$ denote the location of the outermost point measured along $r$ direction on the conformal boundary and brane respectively.
  }\label{fig:QES}
  \end{figure}

In numerical simulations, we take $d=3$, and construct the RT surfaces in the foliation coordinates (\ref{eq:r_theta_y}), which can be described by $r(\theta,y)$, with $0\leq\theta\leq \theta_0$ and $-y_2^*\leq y\leq y_1^*$ -- Fig.~\ref{fig:QES}. 
Substituting $t=0$ and $r=r(\theta,y)$ into (\ref{eq:AdSmetric}), the induced metric on the RT surface reduces to
\begin{equation}\label{eq:induced_metric_on_surface}
    ds^2=\frac{L^2\csc^2\theta}{r^2}\left[ \left((\partial_y r)^2+1\right) dy^2+2  \partial_y r \partial_\theta r \;d\theta  dy+ \left((\partial_\theta r)^2+r^2\right) d\theta^2\right].
\end{equation}
To avoid cumbersome statements, here we have dropped the arguments $(\theta,y)$ in the function $r$. The island formula can be further expressed explicitly as:
\begin{align}\label{eq:QES2}
S[\mathcal{A}]=\lambda \int dy \mathcal{L}_{\text{DGP}}
+\int dy d\theta \mathcal{L}_{\text{bulk}},
\end{align}
where
\begin{align}
    \mathcal{L}_{\text{DGP}}:=&\frac{\csc\theta_0 \sqrt{1+ (\partial_y r)^2}}{r},\\
    \mathcal{L}_{\text{bulk}}:=&\frac{\csc^2\theta \sqrt{\left((\partial_y r)^2+1\right) r^2+(\partial_\theta r)^2}}{r^2}.
\end{align}
For simplicity, in this paper we only consider the cases that possess a mirror symmetry at $y=0$, thus we have $y_1^*=y_2^*$, and can only consider the region $y \geq 0$.
The corresponding four boundary conditions can be imposed as:
\begin{itemize}
    \item \textbf{Dirichlet boundary conditions at} $y=y_*$ :\\
    \begin{equation}
        r(\theta,y_*) = 0.
    \end{equation}
    
    \item \textbf{Symmetric boundary conditions at} $y=0$ :\\
    \begin{equation}
        \partial_y r(\theta,0) = 0.
    \end{equation}
    If one tends to consider a more general case, just replace this condition of Dirichlet boundary conditions at $y=-y_2^*$.
    \item \textbf{Dirichlet boundary conditions at}
    $\theta = 0$ : \\
    \begin{equation}
        r(0,0) = F(y),
    \end{equation}
   {where} $F(y)$ is the shape of $\gamma_A$ on the conformal boundary. The mirror symmetry at $y=0$ further imposes $F(y)=F(-y)$. 
    \item \textbf{Neumann boundary conditions at} $\theta=\theta_0$ : \\
    \begin{equation}\label{eq:NBofQES}
        \partial_\theta r - \lambda \sin\theta_0 \sqrt{(\partial_y r)^2+1}\sqrt{\left((\partial_y r)^2+1\right) r^2+(\partial_\theta r)^2} = 0.
    \end{equation}
    It is worth noting that if we turn off the DGP coupling, the RT surface tends to insert into the brane perpendicularly. Further, (\ref{eq:NBofQES}) also restricts the value of $\lambda$ to be $\lambda\leq \csc\theta_0\leq 1$, which is similar to the discussions in \cite{Ling:2021vxe}.

\end{itemize}
In the following section, we will begin by presenting the universal formulas from both the bulk and brane perspectives{, then} proceed with the numerical analysis based on the general shapes of $\mathcal{A}$ on the conformal boundary from the bulk perspective. Subsequently, we will delve into the specific scenario where $\mathcal{D}$ dominates the contribution of the subregion entanglement.

\section{Entanglement and reflected entropy} \label{sec:3}
In the context of AdS$_4$/BCFT$_3$ duality, when the boundary $\partial \mathcal{A}$ intersects with a flat\footnote{Here, the term ``flat'' refers to the shape of the intersection $\partial \mathcal{B}$ at $x=0$, and not the induced metric on $\mathcal{B}$.} boundary $\partial \mathcal{B}$ at certain corners, the subleading {term} typically exhibits a logarithmic divergence \cite{Seminara:2017hhh,Seminara:2018pmr}.
More recently, in the context of double holography, a dual calculation from the AdS gravity-plus-bath theory has revealed that the entanglement can be decomposed into the contributions from the induced Einstein gravity on the brane, and the terms of the higher curvature together with the quantum fields \cite{Chen:2020uac,Chen:2020hmv,Hernandez:2020nem,Grimaldi:2022suv}. In the following subsection, we will provide a brief review of these two results.
\subsection{The universal behavior}
From the bulk perspective, the entanglement entropy of a finite subregion $\mathcal{W}\cup\mathcal{A}$ in $4$-dimensional spacetime is captured by the formula
\begin{equation}
    S[\mathcal{A}]=\frac{\textbf{Area}(\partial\mathcal{A})}{\epsilon}+\mathcal{F}_{\mathcal{W}\cup\mathcal{A}}+\mathcal{O}(1).
\end{equation}
When $\partial\mathcal{A}$ intersects with the boundary $\partial\mathcal{B}$ at two endpoints that are widely separated compared to the UV cutoff scale, the subleading {term} $\mathcal{F}_{\mathcal{W}\cup\mathcal{A}}$ typically exhibits a logarithmic divergence.
The coefficient preceding the divergence, known as the corner function $\mathcal{G}(\theta_0,\varphi)$, depends on the opening angle $\varphi$ of $\mathcal{A}$ intersecting with $\partial\mathcal{B}$ and the angle $\theta_0$ of the brane intersecting with the conformal boundary $\partial$ \cite{Seminara:2017hhh}.
This leads to the corresponding expression of the subleading divergent term as
\begin{equation}\label{eq:holographic_subleading_term}
  \mathcal{F}_{\mathcal{W}\cup\mathcal{A}}=\mathcal{G}(\theta_0,\varphi) \log \frac{\textbf{Area}(\partial\mathcal{A})}{\epsilon} + \mathcal{O}(1).
\end{equation} 
The characteristics of $\mathcal{G}(\theta_0,\varphi)$ are complicated and depend on both the location of the corners and the boundary conditions that define the BCFT \cite{Seminara:2017hhh,Seminara:2018pmr}.
In particular, for an infinite wedge adjacent to the brane, $\mathcal{G}(\theta_0,\varphi)$ has been derived analytically and subsequently verified numerically for various shapes in finite subregions.

From the brane perspective, in the limit $\theta_0 \to \pi$, the induced action on the brane takes the following form as\cite{Chen:2020uac,Chen:2020hmv,Hernandez:2020nem,Grimaldi:2022suv}
\begin{equation}
  I_{\text{eff}}=I_{b}+I_{\text{reg}},
  \end{equation}
where $I_{b}$ is the terms of the gravitational action (\ref{eq:Action}) on the brane as
\begin{equation}
  I_{b}=
  \frac{-1}{8\pi G_N^{(d+1)}} 
  \int_{\mathcal{B}}d^{d}x\sqrt{-h}\alpha + \frac{1}{16 \pi G_{b}^{(d)}}\int_{\mathcal{B}} d^d x\sqrt{-h}R_{h},
\end{equation}
and $I_{\text{reg}}$ can be expressed as \cite{Emparan:1999pm,Chen:2020uac}
\begin{equation}
  I_{\text{reg}}=\frac{1}{16 \pi G_{N}^{(d+1)}} \int_{\mathcal{B}} d^d x \sqrt{-h}\left[\frac{2(d-1)}{L}+\frac{L}{(d-2)} R_h \right]+\mathcal{O}[R_h]^2.
  \end{equation}
The combination of {these} two actions can be further expressed as
\begin{equation}
  I_{\text{eff}}=\frac{1}{16 \pi G_{\mathrm{eff}}^{(d)}} \int d^d x \sqrt{-h}\left[\frac{(d-1)(d-2)}{\ell_{\mathrm{eff}}^2}+R_h\right]+\mathcal{O}[R_h]^2,
  \end{equation}
with
\begin{equation}\label{eq:effecitve_newton_constant}
  \frac{1}{G_{\mathrm{eff}}^{(d)}}=\left(\frac{1}{d-2}+\lambda\right)\frac{L}{G_{N}^{(d+1)}},\quad\text{and}\quad \frac{1}{l_\text{eff}^2}=\frac{2(d-1-\alpha
   L)}{(d-1)(1+(d-2)\lambda)L^2}.
\end{equation}
In the semi-classical limit, the central charge of the dual conformal defect is given by $c' \sim l_{\text{eff}}^{d-2}/G_{\text{eff}}^{(d)}$, while the central charge of the CFT on this combined system can be given by the standard relation as $c \sim L^{d-1}/G_{N}^{(d+1)}$. Therefore, for $d=3$, the ratio of the central charges can be related to the parameters from the bulk perspective as
\begin{equation}\label{eq:ratio_of_central_charges}
  \frac{c'}{c}\sim (1+\lambda)\sqrt{\frac{1+\lambda}{2+2\cos \theta_0+\lambda \sin^2\theta_0}}.
\end{equation}
It is important to note that (\ref{eq:ratio_of_central_charges}) is only a good approximation at the semi-classical limit with sufficiently small $\theta_0$ or large $\lambda$. (\ref{eq:ratio_of_central_charges}) further indicates that for obtaining a valid ratio, we should have $\lambda>-1$, which actually applies an important lower bound for a proper expansion with respect to the curvature $R_h$ \cite{Chen:2020uac}.
In this scenario, the entanglement entropy (\ref{eq:QES2}) can be re-decomposed as
\begin{equation}\label{eq:QES_New1}
  S[\mathcal{A}]\simeq \bm{A}[\gamma_\mathcal{D}] + S_{c},
\end{equation}
with $\bm{A}[\gamma_\mathcal{D}]:=(1+\lambda)\textbf{Area}(\gamma_\mathcal{D})$.
The first term on the right-hand side, $\bm{A}[\gamma_\mathcal{D}]$, is proportional to the area of QES $\gamma_\mathcal{D}$. This term can be interpreted as the contribution from the induced Einstein gravity on the brane. While the second term, $S_c$, encapsulates all the contributions from high curvature and quantum fields \cite{Chen:2020uac}. In the semiclassical limit, the contributions from high curvature can be effectively suppressed by the ratio $L/\ell_\text{eff}$. Consequently, $S_c$ can be approximated as primarily arising from the entanglement of the quantum fields.

In the following subsections, we will first compare the entanglement behavior from the bulk perspective with that from the brane perspective and analyze their consistency in various scenarios. Then we will elaborate on the effects of the DGP term on the subleading behavior. Finally,
we will exclude the DOF of the bath CFT and investigate the entanglement entropy contributed only by the defect subregion $\mathcal{D}$.

\begin{figure}
  \centering
  \includegraphics[height=0.45\linewidth]{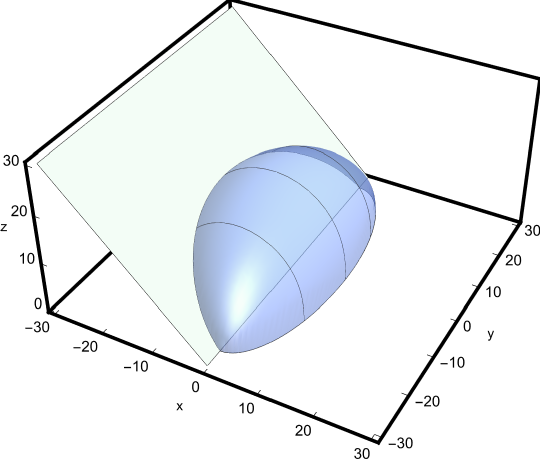}
\caption{A concrete example of the RT surface at $\{\theta_0, \lambda, y^*, a, n, n'\}=\{3\pi/4,0,20,10,2,2\}$.}\label{fig:QESReal}
\end{figure}

\subsection{Numerical Results for semi-ellipses $\mathcal{A}$}
To obtain UV-independent entanglement properties, we set the UV cutoff to $\epsilon=0.05$, which limits the range of admissible parameters. For example, we require $2y^*\gg \epsilon$ in all cases, and $r^*\sin\theta_0 \gg \epsilon$ to capture the logarithmic contributions from the brane. Notably, $a\gg\epsilon$ is not necessary. In fact, the limit $a \to 0$ is significant when we analyze cases of brane dominance, and we will return to this point in the next subsection. Without special notice, we use the tenth-order finite difference method and Chebyshev spectral method in the $\theta$ and $y$ directions, respectively.

\begin{figure}
  \centering
  \subfigure[]{ \label{fig:vary_a_fix_phi}
  \includegraphics[height=0.32\linewidth]{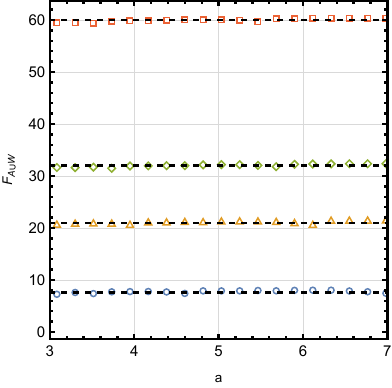}}
  \hspace{0pt}
  \subfigure[]{ \label{fig:vary_ys_fix_phi}
  \includegraphics[height=0.32\linewidth]{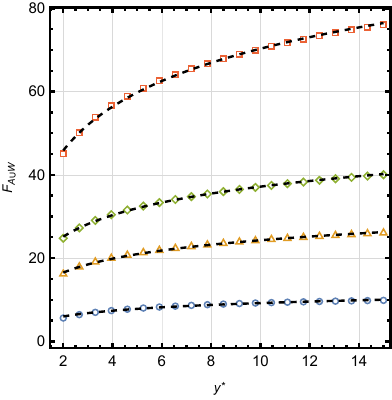}}
  \hspace{0pt}
  \subfigure[]{ \label{fig:c1_fix_phi}
  \includegraphics[height=0.32\linewidth]{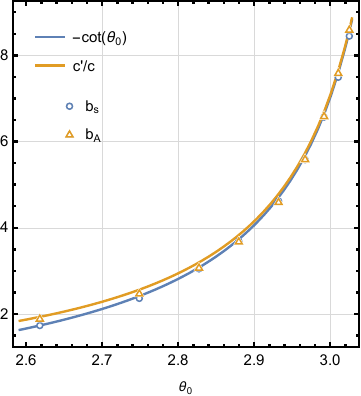}}
\caption{(a): $\mathcal{F}_{\mathcal{W}\cup\mathcal{A}}$ as functions of $a$, with $\{\lambda,y^*,n,n'\}=\{0,5,2,2\}$.(b): $\mathcal{F}_{\mathcal{W}\cup\mathcal{A}}$ as functions of $y^*$, with $\{\lambda,a,n,n'\}=\{0,5,2,2\}$. The blue circles, yellow triangles, green diamonds, and red squares in the first two plots, represent the brane angles with $\theta_0=3\pi/4$, $\theta_0=7\pi/8$, $\theta_0=11\pi/12$ and $\theta_0=23\pi/24$, respectively. With these parameters, we {obtain} solutions with the fixed open angle $\varphi=\pi/2$ at different brane angles $\theta_0$. Hereafter, the black dashed lines represent fitting functions. According to (\ref{eq:FitEntropy}), the fitting parameters $b_s$ are $\simeq1.00$, $\simeq2.41$, $\simeq3.74$ and $\simeq7.59$ respectively for $\theta_0=3\pi/4,\, 7\pi/8,\, 11\pi/12,\, 23\pi/24$. (c): Fitting parameters as functions of $\theta_0$. The blue circles represent the fitting corner functions $b_s$ defined in (\ref{eq:FitEntropy}). While the yellow triangles represent the fitting parameter $b_A$ in $\bm{A}[\gamma_\mathcal{D}]$ when we assume it obeys a similar logarithmic divergence as (\ref{eq:FitEntropy}).}\label{fig:ellipse_fix_phi}
\end{figure}

Consider the shape of the subregion $\mathcal{A}$ on the conformal boundary is described by the following equation
\begin{equation}\label{eq:BoundaryShape}
    F(y)=a \left[1-\frac{y^n}{(y^*)^n}\right]^{1/n'}, \quad F(-y)=F(y), \quad  y^* \geq y \geq 0.
\end{equation}
Here are four adjustable parameters in total $\{a, y^*, n, n'\}$,  with $a>0,\; y\gg \epsilon,\; n>1$ and $n'\geq 1$. 
Refer to Fig.~\ref{fig:QESReal} for a specific construction. For the {subregion with the} shape described by (\ref{eq:BoundaryShape}), the leading linear divergence can be expressed as the following {
integration}:
\begin{equation}\label{eq:leading_divergence}
    \frac{\textbf{Area}(\partial\mathcal{A})}{\epsilon}=\frac{2}{\epsilon} \int_{0}^{y^*} \frac{dy}{n' y}\sqrt{a^2 n^2 \left(\frac{y}{y^*}\right)^{2 n} \left(1-\left(\frac{y}{y^*}\right)^n\right)^{\frac{2}{n'}-2}+n'^2 y^2}.
\end{equation}
For the subleading logarithmic divergence, our numerical simulations suggest the following fitting function:
\begin{equation}\label{eq:FitEntropy}
    \mathcal{F}_{\mathcal{W}\cup\mathcal{A}}= 2 b_s \log \frac{2 y^*}{\epsilon} + b_{s1},
\end{equation}
with $\{b_s,b_{s1}\}$ being two fitting parameters. Note that {
$b_s$} represents the numerically obtained value of the corner function. 

Regarding the secondary divergent term $\mathcal{F}_{\mathcal{W}\cup\mathcal{A}}$, we {have the following remarks based on the illustration} in Fig.~\ref{fig:ellipse_fix_phi}: {First}, the fitting coefficient $b_s$ {is} always equal to $-\cot\theta_0$ corresponding to the special case of the corner function $\mathcal{G}(\theta_0,\pi/2)$ with the open angle {
$\pi/2$} \cite{Seminara:2017hhh}, while the factor $2$ before the corner function represents the {contribution from} two corners of $\mathcal{A}$ on $\partial\mathcal{B}$. Second, the fitting parameter $b_s$ approaches {the ratio of two} central charges (\ref{eq:ratio_of_central_charges}) when $\theta_0 \to \pi$ -- Fig.~\ref{fig:c1_fix_phi}.
At first glance, the similarity of $\mathcal{G}(\theta_0,\varphi)$ and $c'/c$ may be a coincidence, since the corner function varies with the open angle of the extremal surface \cite{Seminara:2017hhh}, while the ratio is determined only by the information of the background spacetime. Therefore, if {two perspectives are consistent}, we {expect that} the corner function $\mathcal{G}(\theta_0,\varphi)$  {would} approach the ratio $c'/c$ for general open angle $\varphi$, as long as $\theta_0$ approaches $\pi$. We will examine this statement with more examples in the next subsection.


Moreover, {we notice} that only the subleading divergent term is influenced by the brane angle, {and} the brane angle is related to the degrees of freedom on the brane. {These facts} encourage us to conjecture that the geometric term from the induced Einstein gravity on brane in (\ref{eq:QES_New1}) contributes a subleading logarithmic divergence, and the prior fitting coefficient $b_A$ can be compared to the ratio of the central charges $c'/c$. As a result, when assuming the geometric term in (\ref{eq:QES_New1}) obeys a logarithmic divergence, the prior coefficient $b_A$ approaches the value of $c'/c$ in the semi-classical limit --Fig.~\ref{fig:c1_fix_phi}. This indicates that the geometric term can be expressed as
\begin{equation}
  \bm{A}[\gamma_\mathcal{D}]\simeq 2 c' \log \frac{\operatorname{Area}(\mathcal{D})}{\epsilon}+\mathcal{O}(1),
\end{equation}
with $c\simeq \frac{L^2}{4G_{N}^{(4)}}=1$, which is valid in the semi-classical limit. 

\begin{figure}
  \centering
    \subfigure[]{\label{fig:rs(y)_at_different_a}
  \includegraphics[height=0.38\linewidth]{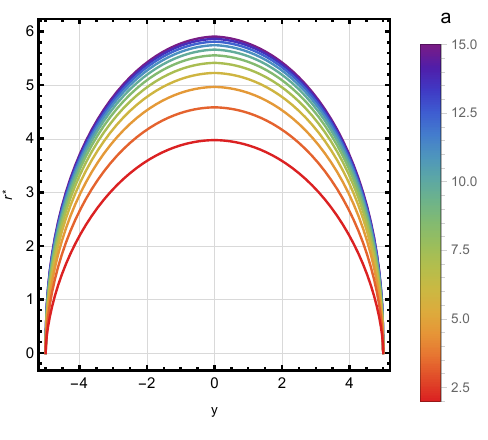}}
  \hspace{0pt}
  \subfigure[]{\label{fig:rs(th0)_at_different_a}
  \includegraphics[height=0.38\linewidth]{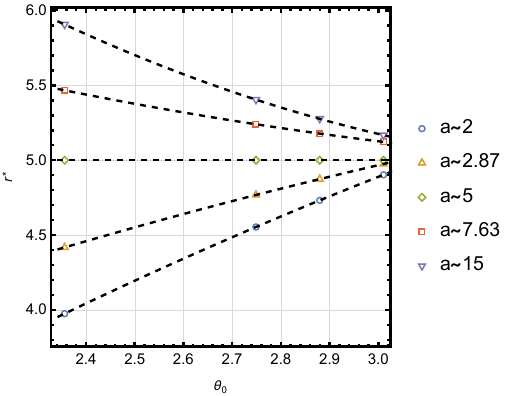}}
\caption{(a): The configurations of $\partial\mathcal{I}$ on the brane with $\{\theta_0, \lambda, y^*, n, n'\}=\{3\pi/4,0,5,2,2\}$. The curves in red and purple represent solutions with $a=2$ and $a=15$, respectively. (b): The locations of the outermost point on the brane $r=r^*$, as functions of the brane angle $\theta_0$, with $\{\lambda, y^*, n, n'\}=\{0,5,2,2\}$. It is obviously that $r^*\simeq y^*$ as $\theta_0 \to \pi$ for all cases.}\label{fig:rs_at_different_a}
\end{figure}

In the context of double holography, the entanglement behavior of the entire conformal defect $\partial\mathcal{B}$ is typically considered when investigating the emergence of quantum extremal islands in brane black holes \cite{Ling:2020laa,Ling:2021vxe,Liu:2022pan}. 
As a result, in the semi-classical limit ($\theta_0 \to \pi$), the corresponding entanglement entropy usually approaches twice the thermodynamical entropy of the black hole, owing to the dominance of $\bm{A}[\gamma_\mathcal{D}]$ \cite{Almheiri:2019yqk,Hashimoto:2020cas,He:2021mst}. 
Nevertheless, in the current scenario with a finite subregion $\mathcal{W}\cup\mathcal{A}$ in the ground state, the entanglement entropy is consistently dominated by $S_c$ for any finite $c'/c$. This phenomenon can be understood from two different perspectives: {First}, from the brane perspective, the linear behavior of the entanglement entropy implies that the entanglement of $\mathcal{W}\cup\mathcal{A}$ is primarily contributed by the quantum fields near $\partial\mathcal{A}$. 
While the logarithmic {behavior of the subleading term} arises from the geometric contribution {of} the induced Einstein gravity on the brane.
Second, from the boundary perspective, the linear behavior can be interpreted as the area law of entanglement for the subsystem $\mathcal{A}$. While the logarithmic {behavior of the subleading term} indicates the entanglement of $\mathcal{D}$ with the remaining part of $\partial \mathcal{B}$.

In summary, the discrepancy of {two} perspectives from general brane angle $\theta_0$ lies in the fact that from the brane perspective \cite{Izumi:2022opi}, the entropy contribution from the subregion on the brane was assumed to be extremized nearly independently of the contribution from the subregion on the conformal boundary. 
As a result, the contribution from the brane always approaches the geodesic length in classical Einstein gravity, as expected from the calculation {in} standard AdS${}_3$ spacetime. 
While from the bulk perspective, we observe that a change in the shape of $\mathcal{A}$ has a significant impact on the configuration of $\mathcal{I}$ on the brane -- Fig.~\ref{fig:rs(y)_at_different_a}. But fortunately, the configuration of $\mathcal{I}$ approaches a semi-circle as $\theta_0 \to \pi$ -- Fig.~\ref{fig:rs(th0)_at_different_a}, which indicates that the solution in the limit of $\theta_0 \to \pi$ approaches the {calculation} obtained from the semi-classical geometry. This result is consistent with \cite{Izumi:2022opi}.

{The above examination is performed with 
} $\mathcal{G}(\theta_0, \varphi=\pi/2)$. More comprehensive examinations will be presented in the subsequent subsection.
A further question arises regarding the measurement of entanglement solely based on $\mathcal{W}$ or, equivalently, the quantum defect $\mathcal{D}$. One direct approach to achieve this is by setting $\varphi \to 0$. In this limit, the entanglement originating from the bath CFT can be excluded, while the entanglement arising from the brane CFT (or the defect) may contribute significantly. We will elaborate on this aspect in the next subsection.

\begin{figure}
  \centering
  \subfigure[]{\label{fig:confiugration_on_the_boundary}
  \includegraphics[height=0.31\linewidth]{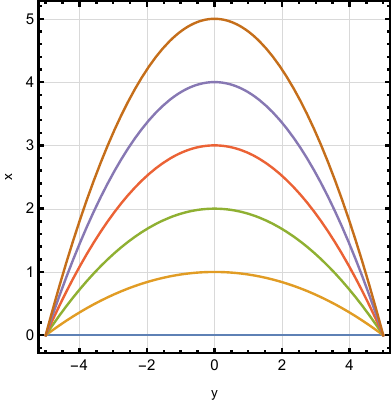}}
  \hspace{0pt}
  \subfigure[]{ \label{fig:rs(a)_at_brane_dominance}
  \includegraphics[height=0.31\linewidth]{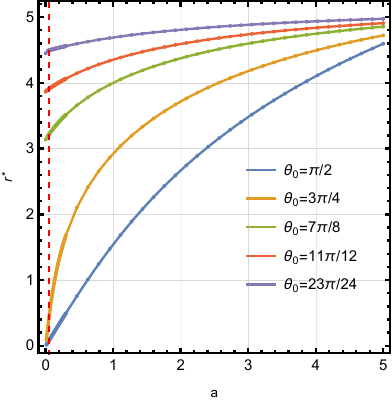}}
  \hspace{0pt}
  \subfigure[]{ \label{fig:FA(a)_at_brane_dominance}
  \includegraphics[height=0.31\linewidth]{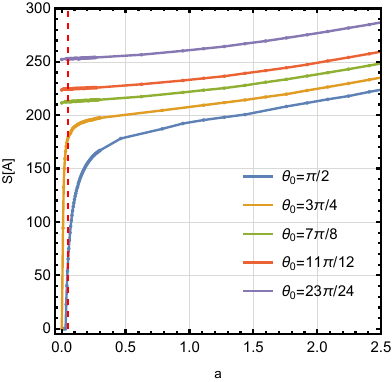}}
\caption{(a): The confiugrations of $\partial\mathcal{A}$ on the conformal boundary, with $\{n,n',y^*\}=\{2,1,5\}$. The curves from the bottom to top are in $a=0,\, 1,\, 2,\, 3,\, 4,\, 5$ respectively. (b): The outermost point on the brane $r^*$ as functions of $a$, (c): the subleading divergent term $\mathcal{F}_{\mathcal{W}\cup\mathcal{A}}$ as functions of $a$. In subfigures (b) and (c), we have $\{\lambda,y^*,n,n'\}=\{0,5,2,1\}$. The red dashed lines in each subfigure correspond to the cutoff scale as $a\simeq\epsilon=0.05$.}
\end{figure}

\subsection{Numerical results for defect subregion $\mathcal{D}$}
Intuitively, the entanglement of $\mathcal{D}$ can be measured by completely removing the bath CFT via $n'=1$, $a \to 0$ and $y^*\gg \epsilon$ -- Fig.~\ref{fig:confiugration_on_the_boundary}. However, the derivative $\partial_\theta r(\theta,y)$ usually diverges in the limit $a \to 0$. Therefore, we should redefine the radial function $R(\theta,y)$ on the extremal surface $\Gamma_{\mathcal{W}\cup\mathcal{A}}$ at $a=0$ as 
\begin{equation}
  r(\theta,y)=\sqrt{(\theta-\frac{\pi}{2})(y^*-y)}R(\theta,y), \quad \text{with} \; \frac{\pi}{2} \leq \theta \leq \theta_0,\; \text{and}\;  0\leq y \leq y^*.
\end{equation}
With this redefinition, $r(\frac{\pi}{2},y)=r(\theta,y^*)=0$ is automatically satisfied. For the solutions right at $a=0$, we use the Chebyshev spectral method in {both} $\theta$ and $y$ directions.

\subsubsection{The phase transition as $a\to0$}
Upon removing the bath CFT, the number of degrees of freedom on the brane will have a substantial impact on the entanglement wedge $\mathcal{W}$ of the defect subregion $\mathcal{D}$ and, consequently, on the entanglement entropy.
In the cases with the lack of degrees of freedom on the brane with small brane angle $\theta_0 \to \pi/2$, we observe a rapid shrinkage of the size of $\mathcal{W}$ as $a \to 0$. Conversely, for cases with $\theta_0 \to \pi$, the presence of an ample number of degrees of freedom on the brane counteracts this shrinkage, thereby maintaining a finite size for $\mathcal{W}$, as illustrated in Fig.~\ref{fig:rs(a)_at_brane_dominance}.
Consequently, the entanglement entropy in each case demonstrates a similar pattern: the subleading divergent term diminishes rapidly for $\theta_0\simeq\pi/2$, while it remains relatively stable for $\theta_0\simeq \pi$, as depicted in Fig.~\ref{fig:FA(a)_at_brane_dominance}. This intriguing phenomenon elicits several noteworthy observations, which we would like to discuss as follows.

First, the illustrations exhibit two distinct phases induced by adjusting the degrees of freedom on the brane. One is the \textbf{shrinking phase}, characterized by $r^*$ or entanglement entropy $S[\mathcal{A}]$ approaching zero as $a\to0$. The other is the \textbf{stable phase}, where $r^*$ remains finite or $\mathcal{F}_{\mathcal{W}\cup \mathcal{A}}$ exhibits relatively stable behavior as $a\to0$.
Thus, there must be a critical angle $\theta_c(\lambda)$ (as a function of $\lambda$) that serves as a threshold to distinguish between these two phases.
Second, the fact that adjusting the degrees of freedom on the brane will lead to a phase transition, is extensively disclosed in the study of double holography, for instance, in \cite{Ling:2020laa,Ling:2021vxe}. In those contexts, increasing the degrees of freedom on the brane leads to a phase transition from the island phase to the trivial phase, which is related to the shrinking and the stable phase in the limit of $y^* \to \infty$.

Last but not least, the rationale behind this phenomenon arises from the area law contribution present in the ground state of the CFT. For positive finite $a$, the entanglement is mainly contributed by the bath CFT near $\partial A$. As removing the bath CFT with $a\to 0$, instead, the entanglement becomes dominated by the brane CFT near $\partial B$, where the spacetime is infinitely stretched.
As a result, the strong entanglement near $\partial B$ leads to the tendency of the entanglement wedge $\mathcal{W}$ to decrease its size as a means of counteraction.
To mitigate this counteractive effect, two potential approaches can be proposed.
The first approach, {as} we have illustrated in Fig.~\ref{fig:FA(a)_at_brane_dominance}, involves enhancing the central charge of the conformal defect $\partial \mathcal{B}$ from {the} boundary perspective \cite{Chen:2020uac,Ling:2020laa,Ling:2021vxe,Liu:2022pan}. For the sufficiently large central charge of the defect $\partial\mathcal{B}$, we have $c'/c\sim (1+\lambda)\csc\theta_0\gg1$. 
This implies that the induced gravity theory on the brane can be approximately {described} by Einstein's theory, and the configuration of $\gamma_\mathcal{D}$ approaches the classical solution.
Specifically, the first approach can be achieved by either increasing the angle $\theta_0$ or DGP coupling $\lambda$. While the second approach involves directly disrupting the area law contribution by raising the temperature of the system, as demonstrated by introducing a black hole \cite{Ling:2021vxe}. This setup can be realized by employing the Einstein-DeTurck formulation to incorporate the backreaction of the Planck brane on the ambient geometry \cite{Headrick:2009pv,Dias:2015nua,Ling:2020laa,Ling:2021vxe,Liu:2022pan}.

In the subsequent discussions, we will focus solely on the first approach, which entails enhancing the central charge of the conformal defect. We will then thoroughly examine the behavior of entanglement in these scenarios.

\begin{figure}
  \centering
  \subfigure[]{ \label{fig:critical_angle_without_lbd}
  \includegraphics[height=0.3\linewidth]{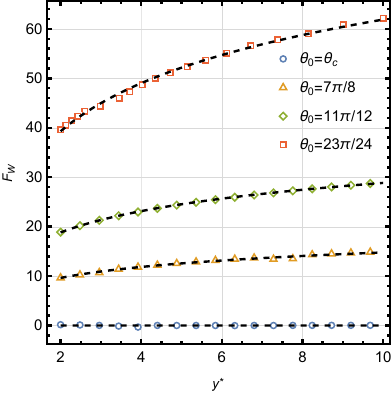}}
  \hspace{0pt}
  \subfigure[]{ \label{fig:critical_angle_against_lbd}
  \includegraphics[height=0.3\linewidth]{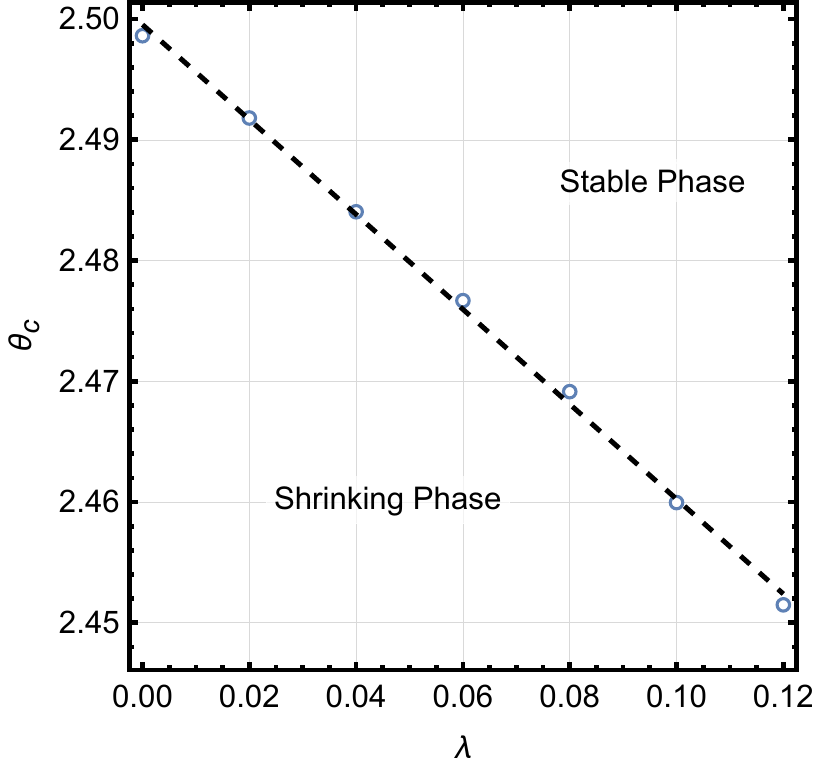}}
  \hspace{0pt}
  \subfigure[]{ \label{fig:c1(lbd)_at_brane_dominance}
  \includegraphics[height=0.3\linewidth]{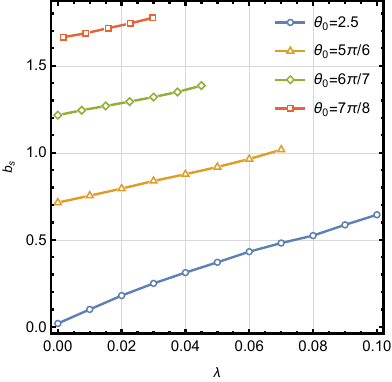}}
\caption{(a): The subleading divergent terms $\mathcal{F}_{\mathcal{W}}$ as functions of $y^*$, with $\{\lambda,a,n,n'\}=\{0,0,2,1\}$ and the critical angle being $\theta_c\simeq 2.5$. The fitting functions are black dashed curves, and the corresponding fitting parameters $b_s$ from the top are $\simeq 7.256$, $\simeq 3.112$, $\simeq 1.601$ and $\simeq 0$, respectively. (b): The critical angles $\theta_c$ at different values of $\lambda$, with $\{a,n,n'\}=\{0,2,1\}$.(c): fitting parameters $b_s$ as functions of $\lambda$, with $\{a,y^*,n,n'\}=\{0,5,2,1\}$.}
\end{figure}
\subsubsection{The entanglement behavior of the defect subsystem $\mathcal{D}$}
\paragraph*{The entanglement entropy of $\mathcal{D}$\\}
Inspired by \cite{Seminara:2017hhh}, the critical brane angle $\theta_0=\theta_c$, which distinguishes the shrinking and stable phases, can be defined when the subleading divergent term $\mathcal{F}_{\mathcal{W}}$ vanishes -- Fig.~\ref{fig:critical_angle_without_lbd}. This indicates that the entanglement entropy of the defect subsystem $\mathcal{D}$ satisfies
\begin{equation}
  S[\mathcal{D}]\simeq\frac{\operatorname{Area}(\mathcal{D})}{\epsilon} +\mathcal{O}(1),
\end{equation}
with $c=1$.
The argument $\mathcal{D}$ instead of $\mathcal{A}$ means that we have set $a\to0$, and the bath subsystem has been entirely removed.
For subcritical brane angles $\theta_0<\theta_c$, both the size of the entanglement wedge $\mathcal{W}$ and the corresponding entanglement entropy decrease rapidly with $a$,
while for supercritical brane angles $\theta_0>\theta_c$, the size of the entanglement wedge $\mathcal{W}$ remains stable -- Fig.~\ref{fig:rs(a)_at_brane_dominance}. 
It is worth noting that, first, the decreasing behavior of entanglement wedge $\mathcal{W}$ for subcritical angles $\theta_0<\theta_c$ resembles the divergence of the corner function $\mathcal{G}(\theta_0,\varphi)$ obtained in the literature \cite{Seminara:2017hhh}, for an infinite wedge as $\theta_0 < \theta_c$ and $\varphi\to0$. 
Second, even after fully excising the contribution from $\mathcal{A}$, the entanglement entropy of the subsystem $\mathcal{D}$ still exhibits a leading linear behavior and a logarithmic subleading behavior.
From the boundary perspective, the linear behavior can be interpreted as a volume law of entanglement for the two-dimensional defect subsystem $\mathcal{D}$. This is due to the fact that, as a boundary, $\mathcal{D}$ directly interacts with the environment, namely, the three-dimensional bath $\bm{\partial}$. On the other hand, the logarithmic subleading behavior indicates its entanglement with the remaining part of $\partial \mathcal{B}$.
From the brane perspective, the linear behavior of the entanglement entropy implies that the entanglement of $\mathcal{W}$ is primarily contributed by the brane CFT near $\partial\mathcal{B}$, where the spacetime is, in fact, infinitely stretched. 
While the subleading logarithmic behavior arises from the geometric contribution.

\begin{figure}
  \centering
  \subfigure[]{ \label{fig:entropy_compared_to_the_generalized_entropy}
  \includegraphics[height=0.39\linewidth]{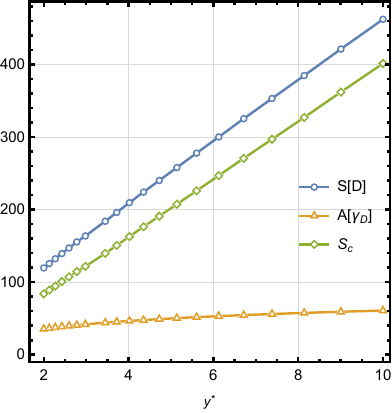}}
  \hspace{50pt}
  \subfigure[]{ \label{fig:coefficient_compared_to_the_generalized_entropy}
  \includegraphics[height=0.39\linewidth]{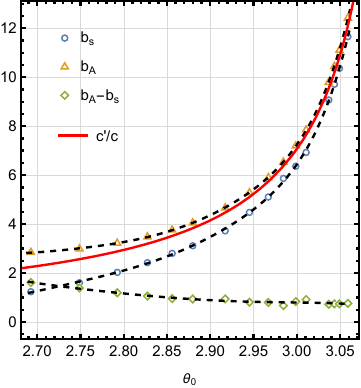}}
\caption{(a): The decomposition of the entanglement entropy $S[\mathcal{D}]$ as functions of $y^*$, with $\{\theta_0,\lambda,a,n,n'\}=\{23\pi/24,0,0,2,1\}$. (b): The comparisons of the fitting parameter $b_s$, with the fitting coefficient $b_A$ before the logarithmic divergence in $\bm{A}[\gamma_\mathcal{D}]$ as functions of the brane angle $\theta_0$. Both subfigures are plotted with the parameters $\{\lambda,a,n,n'\}=\{0,0,2,1\}$.}
\end{figure}

Furthermore, we can consider the influence of the DGP coupling $\lambda$. As illustrated in Fig.~\ref{fig:critical_angle_against_lbd}, the critical angle $\theta_c$ generally decreases with $\lambda$. 
This means that as we increase the DGP coupling $\lambda$, which corresponds to enhancing the degrees of freedom on the brane, it becomes easier for the subsystem $\mathcal{D}$ to undergo a phase transition.
A detailed analysis further reveals that the corner functions $\mathcal{G}$ generally increase with $\lambda$ while keeping the brane angle $\theta_0$ and open-angle $\varphi=0$ fixed -- Fig.~\ref{fig:c1(lbd)_at_brane_dominance}. This implies that the increase in entanglement entropy arising from the DGP term primarily contributes to the subleading divergent term $\mathcal{F}_{\mathcal{W}}$, while the leading behavior, determined solely by the area of $\mathcal{D}$, remains stable.
The physical interpretation behind this observation is that, the variations in $\theta_0$ or $\lambda$ only {affect} on the central charge $c'$. Consequently, the leading divergent term, which reflects the entanglement between the brane and bath CFT, is limited by the central charge $c$ and therefore is not sensitive to these variations.

Following the decomposition suggested in (\ref{eq:QES_New1}), the entanglement entropy of $\mathcal{D}$ is also dominated by $S_c$, which exhibits a linear divergence,  representing the contributions arising from the high curvature and the brane CFT near $\partial\mathcal{B}$  -- Fig.~\ref{fig:entropy_compared_to_the_generalized_entropy}. While the subleading logarithmic divergence is similar to the former cases, representing the contribution from the induced Einstein gravity.  

Prior to the logarithmic divergence, {both coefficients} $b_A$ and $b_s$ approach the ratio $c'/c$ in the semi-classical limit, as depicted in Fig.~\ref{fig:coefficient_compared_to_the_generalized_entropy}. This phenomenon aligns with the previous statement that the entropy primarily originates from the brane CFT near the AdS boundary $\partial \mathcal{B}$, where the spacetime is infinitely stretched.
Therefore, the entanglement entropy of $\mathcal{D}$ in the semi-classical limit suggests the following formula as
\begin{equation}
  S[\mathcal{D}] = c \frac{\operatorname{Area}(\mathcal{D})}{\epsilon} + 2 c'\log \frac{\operatorname{Area}(\mathcal{D})}{\epsilon} + \mathcal{O}(1).
\end{equation}
On the right-hand side, the first term arises from the volume law entanglement between the defect subsystem $\mathcal{D}$ and the bath $\bm{\partial}$, while the second term comes from the logarithmic law entanglement between $\mathcal{D}$ and remaining defect $\partial\mathcal{B}\cap \Bar{\mathcal{D}}$.


\begin{figure}
  \centering
  \includegraphics[height=0.45\linewidth]{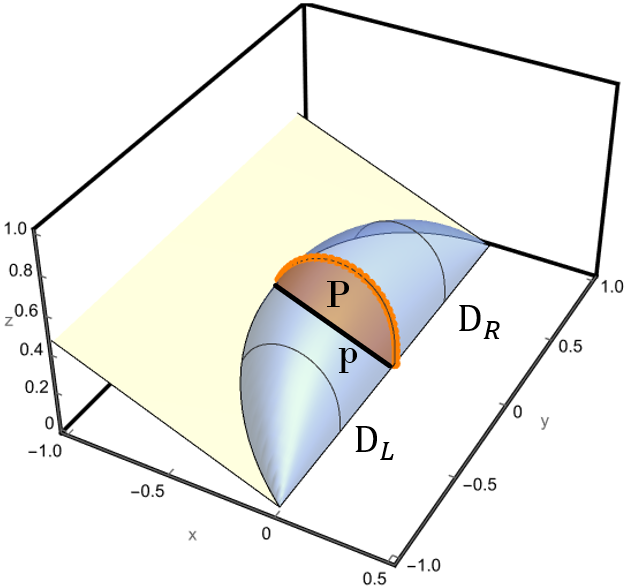}
\caption{A concrete example of the quantum entanglement wedge cross-section $\mathcal{P}$ with the boundary $\bm{p}$ on the brane, at parameters $\{\theta_0, \lambda, y^*, a\}=\{7\pi/8,0,1,0\}$.}\label{fig:QEWCs_Real}
\end{figure}


\paragraph*{The reflected entropy of $\mathcal{D}_{L/R}$\\}
After solving the classical extremal surface $\Gamma_{\mathcal{D}}$, we can further define the reflected entropy via calculating the quantum entanglement wedge cross-section (Q-EWCS) \cite{Ling:2021vxe}.
From the boundary perspective, the reflected entropy {for} two halves of $\mathcal{D}$ with the presence of quantum fields in the bulk, can be defined holographically as
\begin{equation}
  S^R[\mathcal{D}_L:\mathcal{D}_R]=\min_\textbf{p}\left[\frac{\text{Area}(\textbf{p})}{4 G_\text{eff}^{(d)}}+S^R_{c}\right],
\end{equation}
where \textbf{p} is the Q-EWCS separating the entanglement wedge $\mathcal{W}$ on the brane, and $S^R_{c}$ involves the quantum contribution from the brane CFT. While from the brane perspective, $S^R_{c}$ can further be expressed geometrically, and the corresponding formula becomes
\begin{equation}
  S^R[\mathcal{D}_L:\mathcal{D}_R]=\frac{1}{4 G_{N}^{(4)}}\min_\textbf{p}\left[\textbf{Area}(\mathcal{P}) + \lambda L \,\textbf{Area}(\textbf{p})\right],
\end{equation}
where $\mathcal{P}$ is the higher-dimensional EWCS in the bulk, which is bounded by the QEWCS \textbf{p} and the classical extremal surface $\Gamma_\mathcal{D}$. Thanks to the mirror symmetry at $y=0$, the EWCS $\mathcal{P}$ also remains {at} $y=0$ in the current scenario. Refer to Fig.~\ref{fig:QEWCs_Real} for a specific construction.

\begin{figure}
  \centering
  \subfigure[]{ \label{fig:RE(ys)}
  \includegraphics[height=0.3\linewidth]{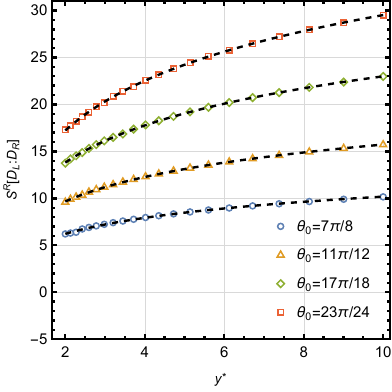}}
  \hspace{0pt}
  \subfigure[]{ \label{fig:partition_of_RE}
  \includegraphics[height=0.3\linewidth]{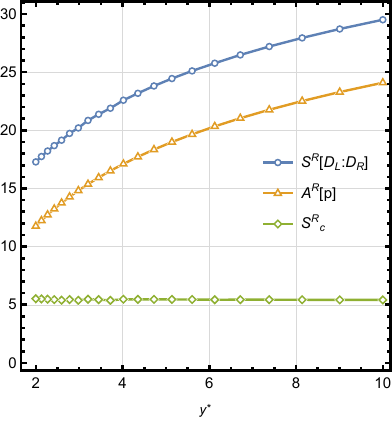}}\\

  \subfigure[]{ \label{fig:corner_function(th0)}
  \includegraphics[height=0.3\linewidth]{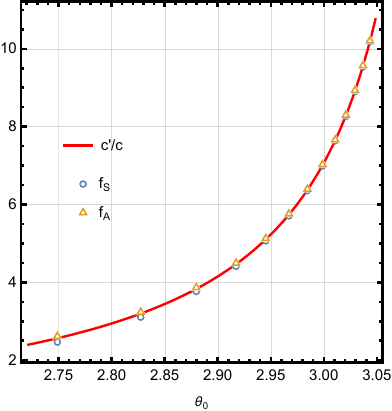}}
  \hspace{0pt}
  \subfigure[]{ \label{fig:correction(th0)}
  \includegraphics[height=0.3\linewidth]{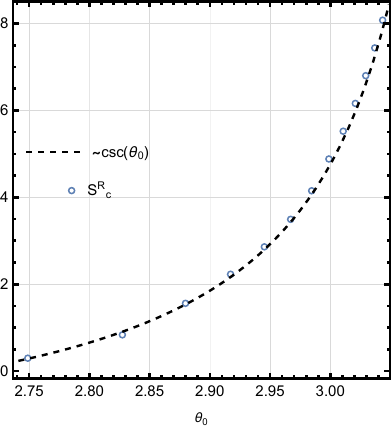}}
 
\caption{(a): The reflected entropy $S^R[\mathcal{D}_L:\mathcal{D}_R]$ as functions of $y^*$, and the brane angle larger than the critical one.
The fitting functions are black-dashed curves, and the corresponding fitting parameters $f_s$ from the top are $\simeq 7.635$, $\simeq 5.707$, $\simeq 3.765$ and $\simeq 2.463$, respectively. (b): The decomposition of the reflected entropy $S^R[\mathcal{D}_L:\mathcal{D}_R]$ with $\theta_0=23\pi/24$. (c): The fitting parameters $f_s$ and $f_A$ prior to the logarithmic divergence of $S^R[\mathcal{D}_L:\mathcal{D}_R]$ and $\operatorname{A}_R[\operatorname{p}]$, and the ratio of central charges $c'/c$ as functions of $\theta_0$, respectively. (d): $S^R_{c}$ as function of $\theta_0$. The fitting function is the black-dashed curve $S^R_{c}\simeq \csc\theta_0-2.33$. All subfigures are plotted with parameters $\{\lambda,a\}=\{0,0\}$.}
\end{figure}

In the stable phase, we observe both entropy $S^R[\mathcal{D}_L:\mathcal{D}_R]$ and geometric contribution $\frac{\text{Area}(\textbf{p})}{4 G_\text{eff}^{(d)}}$ exhibit logarithmic {behavior} with respect to $y^*$, while $S_{Rc}$ remains finite -- Fig.~\ref{fig:RE(ys)} and \ref{fig:partition_of_RE}.  This inspires us to find the numerical fitting functions:
for the reflected entropy and geometric contribution, we impose
\begin{align}
  S^R[\mathcal{D}_L:\mathcal{D}_R]\simeq f_s \log \frac{\operatorname{Area}(\mathcal{D}_{L/R})}{\epsilon}+f_{s_1},\nonumber\\ \frac{\text{Area}(\textbf{p})}{4 G_\text{eff}^{(d)}}\simeq f_A \log \frac{\operatorname{Area}(\mathcal{D}_{L/R})}{\epsilon}+f_{A_1},
\end{align}
with all $f_i\;(i=s,\,s_1,\,A,\,A_1)$ being fitting parameters.
Comparing the fitting parameters with the central charges -- Fig.~\ref{fig:corner_function(th0)} and \ref{fig:correction(th0)}, we observe that the fitting parameters $f_s$ and $f_A$ always approach the ratio $c'/c$, while $S^R_c$ is always a constant up to $\theta_0$.
After recovering the central charge $c$, we finally have the following expression as
\begin{equation}
  S^R[\mathcal{D}_L:\mathcal{D}_R]\simeq c' \log \frac{\operatorname{Area}(\mathcal{D}_{L/R})}{\epsilon} + \text{const.}
\end{equation}
Note that on the right-hand side, the first term arises from the contribution from the induced Einstein gravity on the brane, which always exhibits a logarithmic divergence. While the central charge $c$ has been absorbed {into} the second constant term, and thus the contribution from the quantum fields on the brane is involved in it. Generally speaking, this term does not exhibit any divergence, and is significantly different from the cases in the calculation of entanglement entropy (\ref{eq:QES_New1}).
\section{Conclusions and discussions} \label{sec:4}
In this paper, we {have} investigated the entanglement and reflected entropy of a subregion $\mathcal{W}$ on the brane, which serves as an entanglement wedge of a defect subsystem $\mathcal{D}$. The contribution of the quantum fields {is} studied in the framework of double holography. To compute the entropy holographically, we {have} numerically constructed the corresponding RT surface by solving {the} second-order partial differential {equations} subject to the Neumann boundary condition imposed on the brane. Through this numerical methodology, we {are able to accurately explore the properties of the entanglement in the}  system.

The initial part of our study focused on examining the entanglement entropy of various semi-ellipses $\mathcal{A}$ from the boundary perspective, while also investigating the efficacy of the duality between the brane and boundary viewpoints. Specifically, when the open angle $\varphi$ of the semi-ellipses {is} fixed at $\pi/2$, we {observe} a constant corner function $\mathcal{G}(\theta_0,\varphi)$, which {appears in the subleading term of} logarithmic divergence, across different values of the brane angle $\theta_0$. This finding aligns with previous predictions reported in the literature \cite{Seminara:2017hhh}.
In the semiclassical limit, where $c'/c$ approached infinity, we {have observed} a consistent attraction of the boundary contour of the island, denoted as $\partial\mathcal{I}$, towards a semi-circle for different semi-ellipses. As a result, the geometric contribution originating from the induced Einstein gravity on the brane {approaches} the classical solution. This outcome aligns with the semiclassical framework presented from the brane perspective, adding further coherence to our analysis.

The second part of our {investigation primarily focuses on} the entanglement behavior of the defect subsystem $\mathcal{D}$.  Without a sufficient amount of degrees of freedom on the brane, the removal of the contribution from the bath CFT in $\mathcal{A}$ may result in the shrinkage of the size of $\mathcal{W}$ due to the counteraction of its strong entanglement with the bath. We refer to this as the shrinking phase, characterized by a rapid reduction in entropy with the removal of $\mathcal{A}$. However, when the number of degrees of freedom on the brane is increased, a phase transition occurs, counteracting the shrinkage effect. We refer to this as the stable phase, where the size of $\mathcal{W}$ and the corresponding entropy remain nearly constant. It is important to note that the semiclassical limit {is} defined within this specific parameter regime.

On {one} side, we {have} investigated the entanglement entropy of the defect subsystem $\mathcal{D}$. In the semiclassical limit, the geometric contribution of the area of the quantum extremal surface also gives rise to a subleading logarithmic divergence. Meanwhile, the quantum fields within $\mathcal{W}$ on the brane contribute a leading linear divergence. 
From the brane perspective, the linear behavior indicates that the entanglement of the brane CFT within $\mathcal{W}$ is primarily influenced by the quantum fields near the AdS boundary $\partial\mathcal{B}$, where the spacetime is infinitely stretched. 
While the subleading logarithmic behavior arises from the geometric contribution of the induced Einstein gravity on the brane.
From the boundary perspective, the linear behavior can be interpreted as a volume law of entanglement. This is due to the direct interaction of $\mathcal{D}$ with the higher-dimensional bath $\bm{\partial}$. 
The subleading logarithmic behavior, in contrast, suggests its entanglement with the remaining part of $\partial \mathcal{B}$.

Moreover, it is important to note that the DGP coupling also affects the degrees of freedom on the brane. Consequently, an increase in the DGP coupling primarily impacts {on} the central charge of the brane and bath CFT, which appears {in the subleading term with} logarithmic divergence. Furthermore, since the phase transition is solely induced by the degrees of freedom on the brane, an increase in the DGP coupling makes it easier for the subsystem $\mathcal{W}$ to enter the regime of the stable phase. This is reflected in the decrease of the critical brane angle. In other words, a larger DGP coupling allows for a wider range of parameter values for which the stable phase can be observed.

On the other side, we {have} also investigated the reflected entropy within the defect subsystem $\mathcal{D}$. Specifically, by dividing $\mathcal{D}$ into two equal-sized subsystems, $\mathcal{D}_L$ and $\mathcal{D}_R$, the reflected entropy {is} calculated using the Q-EWCS that separates the entanglement wedge $\mathcal{W}$. The contribution of the brane CFT within $\mathcal{W}$ to the reflected entropy {is} further analyzed within the framework of double holography. In the semiclassical limit, it {is} observed that the entanglement from the brane CFT within $\mathcal{D}$ only {provides} a finite term contribution, while the dominant contribution consistently {arises} from the induced Einstein gravity on the brane.

It is also interesting to investigate the mutual information between two disjoint subsystems, denoted as $\mathcal{D}_1$ and $\mathcal{D}_2$, on the conformal defect $\partial \mathcal{B}$. In the context of {standard} AdS${}_3$/CFT${}_2$ correspondence, the extremal surface of the composite system $\mathcal{D}_1\cup\mathcal{D}_2$ exhibits two distinct phases: the connected phase, bounded by two semi-circles with the one inside the other, and the disconnected phase bounded by two separate semi-circles. It is well known that the holographic mutual information vanishes at the leading order when the extremal surface is in the disconnected phase. 
On the contrary, in the AdS-gravity-plus-bath scenario, an additional intermediate phase may arise. 
Even when the extremal surface on the brane, denoted as $\mathcal{W}$, is in the disconnected phase, the holographic mutual information between $\mathcal{D}_1$ and $\mathcal{D}_2$ can still be positive. This positivity arises from the supplementary entanglement contributed by the quantum fields {presented} in the background and indicates a connected entanglement wedge in the framework of double holography.

Another intriguing aspect to explore is the effect of increasing the temperature in the current system, which can be achieved by introducing a black hole. Technically, the gravitational background can be solved via the Einstein-DeTurck formulation \cite{Headrick:2009pv,Dias:2015nua}. As mentioned earlier, apart from increasing the ratio of central charges, raising the temperature of the system is another approach to disrupt long-range correlations. Consequently, in this scenario, we anticipate that the system will consistently remain in the stable phase, without undergoing a phase transition induced by the increased degrees of freedom on the brane. Additionally, we also expect the subleading behavior of the entanglement entropy to deviate from a logarithmic divergence and exhibit a linear divergence instead. This linear divergence arises from the volume law of entanglement when the system's scale approaches the temperature scale.

The ground state and the thermal state of the system can be further connected by a holographic quench. In terms of the brane perspective, this far-from-equilibrium process corresponds to the formation of brane black holes from the vacuum state. As a result, the entanglement production in the gravitational region can be investigated during this thermalization process \cite{Janik:2006ft,Chesler:2008hg,Bhattacharyya:2009uu}. 
Additionally, the entanglement production of more general far-from-equilibrium processes can also be further studied in the future.
For a spherically symmetric gravitational dynamical system, an unstable Reissner-Nordstr$\ddot{\text{o}}$m-AdS black hole will dynamically scalarize under arbitrary perturbation, leading to the spontaneous scalarization or superradiance, which depends on whether the scalar field is charged or not \cite{Bosch:2016vcp,Bosch:2019anc,Zhang:2021etr,Zhang:2022cmu,Chen:2022vag,Chen:2023eru}.
For a planar black hole, on the other hand, a thermodynamically favored inhomogeneous black hole can be dynamically generated from a homogeneous black hole with a spinodal instability, resulting in the emergence of phase separation \cite{Janik:2017ykj,Chen:2022cwi,Chen:2022tfy}.

\section*{Acknowledgments}
We are grateful to Yuan Sun for the helpful discussions. Liu Yuxuan special thanks to Peiwen Cao for supporting his work. This work is supported by China Postdoctoral Science Foundation, under the National Postdoctoral Program for Innovative Talents BX2021303. {Y.Ling is supported in part by the Natural Science Foundation of China under Grant No.~12035016 and 12275275. It is also supported by Beijing Natural Science Foundation under Grant No. 1222031, and the Innovative Projects of Science and Technology at IHEP.} CP is supported by NSFC NO.~12175237, the Fundamental Research Funds for the Central Universities, and funds from the Chinese Academy of Sciences. ZYX is funded by DFG through the Collaborative Research Center SFB 1170 ToCoTronics, Project-ID 258499086—SFB 1170, as well as by Germany's Excellence Strategy through the W\"urzburg‐Dresden Cluster of Excellence on Complexity and Topology in Quantum Matter ‐ ct.qmat (EXC 2147, project‐id 390858490). ZYX also acknowledges support from the National Natural Science Foundation of China under Grants No.~11875053 and No.~12075298.

\bibliographystyle{unsrt}

\bibliography{ref}

\begin{thebibliography}{100}

\bibitem{Maldacena:1997re}
Juan~Martin Maldacena.
\newblock {The Large N limit of superconformal field theories and
  supergravity}.
\newblock {\em Adv. Theor. Math. Phys.}, 2:231--252, 1998.

\bibitem{Gubser:1998bc}
S.~S. Gubser, Igor~R. Klebanov, and Alexander~M. Polyakov.
\newblock {Gauge theory correlators from noncritical string theory}.
\newblock {\em Phys. Lett. B}, 428:105--114, 1998.

\bibitem{Witten:1998qj}
Edward Witten.
\newblock {Anti-de Sitter space and holography}.
\newblock {\em Adv. Theor. Math. Phys.}, 2:253--291, 1998.

\bibitem{Ryu:2006bv}
Shinsei Ryu and Tadashi Takayanagi.
\newblock {Holographic derivation of entanglement entropy from AdS/CFT}.
\newblock {\em Phys. Rev. Lett.}, 96:181602, 2006.

\bibitem{Ryu:2006ef}
Shinsei Ryu and Tadashi Takayanagi.
\newblock {Aspects of Holographic Entanglement Entropy}.
\newblock {\em JHEP}, 08:045, 2006.

\bibitem{Hubeny:2007xt}
Veronika~E. Hubeny, Mukund Rangamani, and Tadashi Takayanagi.
\newblock {A Covariant holographic entanglement entropy proposal}.
\newblock {\em JHEP}, 07:062, 2007.

\bibitem{Lewkowycz:2013nqa}
Aitor Lewkowycz and Juan Maldacena.
\newblock {Generalized gravitational entropy}.
\newblock {\em JHEP}, 08:090, 2013.

\bibitem{Engelhardt:2014gca}
Netta Engelhardt and Aron~C. Wall.
\newblock {Quantum Extremal Surfaces: Holographic Entanglement Entropy beyond
  the Classical Regime}.
\newblock {\em JHEP}, 01:073, 2015.

\bibitem{Witten:2021unn}
Edward Witten.
\newblock {Gravity and the crossed product}.
\newblock {\em JHEP}, 10:008, 2022.

\bibitem{Almheiri:2019hni}
Ahmed Almheiri, Raghu Mahajan, Juan Maldacena, and Ying Zhao.
\newblock {The Page curve of Hawking radiation from semiclassical geometry}.
\newblock {\em JHEP}, 03:149, 2020.

\bibitem{Almheiri:2019psf}
Ahmed Almheiri, Netta Engelhardt, Donald Marolf, and Henry Maxfield.
\newblock {The entropy of bulk quantum fields and the entanglement wedge of an
  evaporating black hole}.
\newblock {\em JHEP}, 12:063, 2019.

\bibitem{Almheiri:2019yqk}
Ahmed Almheiri, Raghu Mahajan, and Juan Maldacena.
\newblock {Islands outside the horizon}.
\newblock {\em arXiv:1910.11077}, 10 2019.

\bibitem{Penington:2019kki}
Geoff Penington, Stephen~H. Shenker, Douglas Stanford, and Zhenbin Yang.
\newblock {Replica wormholes and the black hole interior}.
\newblock {\em JHEP}, 03:205, 2022.

\bibitem{Almheiri:2019qdq}
Ahmed Almheiri, Thomas Hartman, Juan Maldacena, Edgar Shaghoulian, and
  Amirhossein Tajdini.
\newblock {Replica Wormholes and the Entropy of Hawking Radiation}.
\newblock {\em JHEP}, 05:013, 2020.

\bibitem{Almheiri:2019psy}
Ahmed Almheiri, Raghu Mahajan, and Jorge~E. Santos.
\newblock {Entanglement islands in higher dimensions}.
\newblock {\em SciPost Phys.}, 9(1):001, 2020.

\bibitem{Chen:2019uhq}
Hong~Zhe Chen, Zachary Fisher, Juan Hernandez, Robert~C. Myers, and Shan-Ming
  Ruan.
\newblock {Information Flow in Black Hole Evaporation}.
\newblock {\em JHEP}, 03:152, 2020.

\bibitem{Chen:2020uac}
Hong~Zhe Chen, Robert~C. Myers, Dominik Neuenfeld, Ignacio~A. Reyes, and Joshua
  Sandor.
\newblock {Quantum Extremal Islands Made Easy, Part I: Entanglement on the
  Brane}.
\newblock {\em JHEP}, 10:166, 2020.

\bibitem{Chen:2020hmv}
Hong~Zhe Chen, Robert~C. Myers, Dominik Neuenfeld, Ignacio~A. Reyes, and Joshua
  Sandor.
\newblock {Quantum Extremal Islands Made Easy, Part II: Black Holes on the
  Brane}.
\newblock {\em JHEP}, 12:025, 2020.

\bibitem{Hernandez:2020nem}
Juan Hernandez, Robert~C. Myers, and Shan-Ming Ruan.
\newblock {Quantum extremal islands made easy. Part III. Complexity on the
  brane}.
\newblock {\em JHEP}, 02:173, 2021.

\bibitem{Grimaldi:2022suv}
Guglielmo Grimaldi, Juan Hernandez, and Robert~C. Myers.
\newblock {Quantum extremal islands made easy. Part IV. Massive black holes on
  the brane}.
\newblock {\em JHEP}, 03:136, 2022.

\bibitem{hawking1974black}
Stephen~W Hawking.
\newblock Black hole explosions?
\newblock {\em Nature}, 248(5443):30--31, 1974.

\bibitem{hawking1975particle}
Stephen~W Hawking.
\newblock Particle creation by black holes.
\newblock In {\em Euclidean quantum gravity}, pages 167--188. World Scientific,
  1975.

\bibitem{hawking1976breakdown}
Stephen~W Hawking.
\newblock Breakdown of predictability in gravitational collapse.
\newblock {\em Physical Review D}, 14(10):2460, 1976.

\bibitem{Page:1993wv}
Don~N. Page.
\newblock {Information in black hole radiation}.
\newblock {\em Phys. Rev. Lett.}, 71:3743--3746, 1993.

\bibitem{susskind1993stretched}
Leonard Susskind, Larus Thorlacius, and John Uglum.
\newblock The stretched horizon and black hole complementarity.
\newblock {\em Physical Review D}, 48(8):3743, 1993.

\bibitem{Page:2004xp}
Don~N. Page.
\newblock {Hawking radiation and black hole thermodynamics}.
\newblock {\em New J. Phys.}, 7:203, 2005.

\bibitem{Almheiri:2012rt}
Ahmed Almheiri, Donald Marolf, Joseph Polchinski, and James Sully.
\newblock {Black Holes: Complementarity or Firewalls?}
\newblock {\em JHEP}, 02:062, 2013.

\bibitem{Page:2013dx}
Don~N. Page.
\newblock {Time Dependence of Hawking Radiation Entropy}.
\newblock {\em JCAP}, 09:028, 2013.

\bibitem{Hartman:2013qma}
Thomas Hartman and Juan Maldacena.
\newblock {Time Evolution of Entanglement Entropy from Black Hole Interiors}.
\newblock {\em JHEP}, 05:014, 2013.

\bibitem{Takayanagi:2011zk}
Tadashi Takayanagi.
\newblock {Holographic Dual of BCFT}.
\newblock {\em Phys. Rev. Lett.}, 107:101602, 2011.

\bibitem{Chu:2018ntx}
Chong-Sun Chu and Rong-Xin Miao.
\newblock {Anomalous Transport in Holographic Boundary Conformal Field
  Theories}.
\newblock {\em JHEP}, 07:005, 2018.

\bibitem{Miao:2018qkc}
Rong-Xin Miao.
\newblock {Holographic BCFT with Dirichlet Boundary Condition}.
\newblock {\em JHEP}, 02:025, 2019.

\bibitem{Ling:2020laa}
Yi~Ling, Yuxuan Liu, and Zhuo-Yu Xian.
\newblock {Island in Charged Black Holes}.
\newblock {\em JHEP}, 03:251, 2021.

\bibitem{Geng:2020fxl}
Hao Geng, Andreas Karch, Carlos Perez-Pardavila, Suvrat Raju, Lisa Randall,
  Marcos Riojas, and Sanjit Shashi.
\newblock {Information Transfer with a Gravitating Bath}.
\newblock {\em SciPost Phys.}, 10(5):103, 2021.

\bibitem{Geng:2020qvw}
Hao Geng and Andreas Karch.
\newblock {Massive islands}.
\newblock {\em JHEP}, 09:121, 2020.

\bibitem{Krishnan:2020fer}
Chethan Krishnan.
\newblock {Critical Islands}.
\newblock {\em JHEP}, 01:179, 2021.

\bibitem{Miao:2020oey}
Rong-Xin Miao.
\newblock {An Exact Construction of Codimension two Holography}.
\newblock {\em JHEP}, 01:150, 2021.

\bibitem{Akal:2020wfl}
Ibrahim Akal, Yuya Kusuki, Tadashi Takayanagi, and Zixia Wei.
\newblock {Codimension two holography for wedges}.
\newblock {\em Phys. Rev. D}, 102(12):126007, 2020.

\bibitem{Akal:2020twv}
Ibrahim Akal, Yuya Kusuki, Noburo Shiba, Tadashi Takayanagi, and Zixia Wei.
\newblock {Entanglement Entropy in a Holographic Moving Mirror and the Page
  Curve}.
\newblock {\em Phys. Rev. Lett.}, 126(6):061604, 2021.

\bibitem{Omidi:2021opl}
Farzad Omidi.
\newblock {Entropy of Hawking radiation for two-sided hyperscaling violating
  black branes}.
\newblock {\em JHEP}, 04:022, 2022.

\bibitem{Rozali:2019day}
Moshe Rozali, James Sully, Mark Van~Raamsdonk, Christopher Waddell, and David
  Wakeham.
\newblock {Information radiation in BCFT models of black holes}.
\newblock {\em JHEP}, 05:004, 2020.

\bibitem{Karlsson:2020uga}
Anna Karlsson.
\newblock {Replica wormhole and island incompatibility with monogamy of
  entanglement}.
\newblock 7 2020.

\bibitem{Balasubramanian:2020xqf}
Vijay Balasubramanian, Arjun Kar, and Tomonori Ugajin.
\newblock {Islands in de Sitter space}.
\newblock {\em JHEP}, 02:072, 2021.

\bibitem{Balasubramanian:2021wgd}
Vijay Balasubramanian, Arjun Kar, and Tomonori Ugajin.
\newblock {Entanglement between two gravitating universes}.
\newblock {\em Class. Quant. Grav.}, 39(17):174001, 2022.

\bibitem{Balasubramanian:2020coy}
Vijay Balasubramanian, Arjun Kar, and Tomonori Ugajin.
\newblock {Entanglement between two disjoint universes}.
\newblock {\em JHEP}, 02:136, 2021.

\bibitem{Miyata:2021ncm}
Akihiro Miyata and Tomonori Ugajin.
\newblock {Evaporation of black holes in flat space entangled with an auxiliary
  universe}.
\newblock {\em PTEP}, 2022(1):013B13, 2022.

\bibitem{Miyata:2021qsm}
Akihiro Miyata and Tomonori Ugajin.
\newblock {Entanglement between two evaporating black holes}.
\newblock {\em JHEP}, 09:009, 2022.

\bibitem{Marolf:2020rpm}
Donald Marolf and Henry Maxfield.
\newblock {Observations of Hawking radiation: the Page curve and baby
  universes}.
\newblock {\em JHEP}, 04:272, 2021.

\bibitem{Marolf:2020xie}
Donald Marolf and Henry Maxfield.
\newblock {Transcending the ensemble: baby universes, spacetime wormholes, and
  the order and disorder of black hole information}.
\newblock {\em JHEP}, 08:044, 2020.

\bibitem{Balasubramanian:2020jhl}
Vijay Balasubramanian, Arjun Kar, Simon~F. Ross, and Tomonori Ugajin.
\newblock {Spin structures and baby universes}.
\newblock {\em JHEP}, 09:192, 2020.

\bibitem{Peng:2021vhs}
Cheng Peng, Jia Tian, and Jianghui Yu.
\newblock {Baby universes, ensemble averages and factorizations with matters}.
\newblock {\em arXiv preprint arXiv:2111.14856}, 11 2021.

\bibitem{Peng:2022pfa}
Cheng Peng, Jia Tian, and Yingyu Yang.
\newblock {Half-wormholes and ensemble averages}.
\newblock {\em Eur. Phys. J. C}, 83(11):993, 2023.

\bibitem{Alishahiha:2020qza}
Mohsen Alishahiha, Amin Faraji~Astaneh, and Ali Naseh.
\newblock {Island in the presence of higher derivative terms}.
\newblock {\em JHEP}, 02:035, 2021.

\bibitem{Hashimoto:2020cas}
Koji Hashimoto, Norihiro Iizuka, and Yoshinori Matsuo.
\newblock {Islands in Schwarzschild black holes}.
\newblock {\em JHEP}, 06:085, 2020.

\bibitem{Anegawa:2020ezn}
Takanori Anegawa and Norihiro Iizuka.
\newblock {Notes on islands in asymptotically flat 2d dilaton black holes}.
\newblock {\em JHEP}, 07:036, 2020.

\bibitem{Hartman:2020swn}
Thomas Hartman, Edgar Shaghoulian, and Andrew Strominger.
\newblock {Islands in Asymptotically Flat 2D Gravity}.
\newblock {\em JHEP}, 07:022, 2020.

\bibitem{Chen:2020jvn}
Hong~Zhe Chen, Zachary Fisher, Juan Hernandez, Robert~C. Myers, and Shan-Ming
  Ruan.
\newblock {Evaporating Black Holes Coupled to a Thermal Bath}.
\newblock {\em JHEP}, 01:065, 2021.

\bibitem{Bhattacharya:2020uun}
Aranya Bhattacharya, Anindya Chanda, Sabyasachi Maulik, Christian Northe, and
  Shibaji Roy.
\newblock {Topological shadows and complexity of islands in multiboundary
  wormholes}.
\newblock {\em JHEP}, 02:152, 2021.

\bibitem{Deng:2020ent}
Feiyu Deng, Jinwei Chu, and Yang Zhou.
\newblock {Defect extremal surface as the holographic counterpart of Island
  formula}.
\newblock {\em JHEP}, 03:008, 2021.

\bibitem{Wang:2021woy}
Xuanhua Wang, Ran Li, and Jin Wang.
\newblock {Islands and Page curves of Reissner-Nordstr\"om black holes}.
\newblock {\em JHEP}, 04:103, 2021.

\bibitem{He:2021mst}
Song He, Yuan Sun, Long Zhao, and Yu-Xuan Zhang.
\newblock {The universality of islands outside the horizon}.
\newblock {\em JHEP}, 05:047, 2022.

\bibitem{Gautason:2020tmk}
F.~F. Gautason, Lukas Schneiderbauer, Watse Sybesma, and L\'arus Thorlacius.
\newblock {Page Curve for an Evaporating Black Hole}.
\newblock {\em JHEP}, 05:091, 2020.

\bibitem{Krishnan:2020oun}
Chethan Krishnan, Vaishnavi Patil, and Jude Pereira.
\newblock {Page Curve and the Information Paradox in Flat Space}.
\newblock 5 2020.

\bibitem{Sybesma:2020fxg}
Watse Sybesma.
\newblock {Pure de Sitter space and the island moving back in time}.
\newblock {\em Class. Quant. Grav.}, 38(14):145012, 2021.

\bibitem{Chou:2021boq}
Chia-Jui Chou, Hans~B. Lao, and Yi~Yang.
\newblock {Page curve of effective Hawking radiation}.
\newblock {\em Phys. Rev. D}, 106(6):066008, 2022.

\bibitem{Hollowood:2021lsw}
Timothy~J. Hollowood, S.~Prem Kumar, Andrea Legramandi, and Neil Talwar.
\newblock {Grey-body factors, irreversibility and multiple island saddles}.
\newblock {\em JHEP}, 03:110, 2022.

\bibitem{Suzuki:2022xwv}
Kenta Suzuki and Tadashi Takayanagi.
\newblock {BCFT and Islands in two dimensions}.
\newblock {\em JHEP}, 06:095, 2022.

\bibitem{Suzuki:2022yru}
Yu-ki Suzuki and Seiji Terashima.
\newblock {On the dynamics in the AdS/BCFT correspondence}.
\newblock {\em JHEP}, 09:103, 2022.

\bibitem{Bhattacharya:2021nqj}
Aranya Bhattacharya, Arpan Bhattacharyya, Pratik Nandy, and Ayan~K. Patra.
\newblock {Bath deformations, islands, and holographic complexity}.
\newblock {\em Phys. Rev. D}, 105(6):066019, 2022.

\bibitem{Bhattacharya:2021dnd}
Aranya Bhattacharya, Arpan Bhattacharyya, Pratik Nandy, and Ayan~K. Patra.
\newblock {Partial islands and subregion complexity in geometric secret-sharing
  model}.
\newblock {\em JHEP}, 12:091, 2021.

\bibitem{Caceres:2021fuw}
Elena Caceres, Arnab Kundu, Ayan~K. Patra, and Sanjit Shashi.
\newblock {Page curves and bath deformations}.
\newblock {\em SciPost Phys. Core}, 5:033, 2022.

\bibitem{Bhattacharya:2021jrn}
Aranya Bhattacharya, Arpan Bhattacharyya, Pratik Nandy, and Ayan~K. Patra.
\newblock {Islands and complexity of eternal black hole and radiation
  subsystems for a doubly holographic model}.
\newblock {\em JHEP}, 05:135, 2021.

\bibitem{Caceres:2020jcn}
Elena Caceres, Arnab Kundu, Ayan~K. Patra, and Sanjit Shashi.
\newblock {Warped information and entanglement islands in AdS/WCFT}.
\newblock {\em JHEP}, 07:004, 2021.

\bibitem{Chen:2019iro}
Yiming Chen.
\newblock {Pulling Out the Island with Modular Flow}.
\newblock {\em JHEP}, 03:033, 2020.

\bibitem{Balasubramanian:2020hfs}
Vijay Balasubramanian, Arjun Kar, Onkar Parrikar, G\'abor S\'arosi, and
  Tomonori Ugajin.
\newblock {Geometric secret sharing in a model of Hawking radiation}.
\newblock {\em JHEP}, 01:177, 2021.

\bibitem{Almheiri:2020cfm}
Ahmed Almheiri, Thomas Hartman, Juan Maldacena, Edgar Shaghoulian, and
  Amirhossein Tajdini.
\newblock {The entropy of Hawking radiation}.
\newblock {\em Rev. Mod. Phys.}, 93(3):035002, 2021.

\bibitem{Li:2020ceg}
Tianyi Li, Jinwei Chu, and Yang Zhou.
\newblock {Reflected Entropy for an Evaporating Black Hole}.
\newblock {\em JHEP}, 11:155, 2020.

\bibitem{KumarBasak:2020ams}
Jaydeep Kumar~Basak, Debarshi Basu, Vinay Malvimat, Himanshu Parihar, and
  Gautam Sengupta.
\newblock {Islands for entanglement negativity}.
\newblock {\em SciPost Phys.}, 12(1):003, 2022.

\bibitem{Anderson:2020vwi}
Louise Anderson, Onkar Parrikar, and Ronak~M. Soni.
\newblock {Islands with gravitating baths: towards ER = EPR}.
\newblock {\em JHEP}, 21:226, 2020.

\bibitem{Vardhan:2021mdy}
Shreya Vardhan, Jonah Kudler-Flam, Hassan Shapourian, and Hong Liu.
\newblock {Mixed-state entanglement and information recovery in thermalized
  states and evaporating black holes}.
\newblock {\em JHEP}, 01:064, 2023.

\bibitem{Kawabata:2021vyo}
Kohki Kawabata, Tatsuma Nishioka, Yoshitaka Okuyama, and Kento Watanabe.
\newblock {Replica wormholes and capacity of entanglement}.
\newblock {\em JHEP}, 10:227, 2021.

\bibitem{Kawabata:2021hac}
Kohki Kawabata, Tatsuma Nishioka, Yoshitaka Okuyama, and Kento Watanabe.
\newblock {Probing Hawking radiation through capacity of entanglement}.
\newblock {\em JHEP}, 05:062, 2021.

\bibitem{Geng:2021iyq}
Hao Geng, Severin L\"ust, Rashmish~K. Mishra, and David Wakeham.
\newblock {Holographic BCFTs and Communicating Black Holes}.
\newblock {\em jhep}, 08:003, 2021.

\bibitem{Geng:2021mic}
Hao Geng, Andreas Karch, Carlos Perez-Pardavila, Suvrat Raju, Lisa Randall,
  Marcos Riojas, and Sanjit Shashi.
\newblock {Entanglement phase structure of a holographic BCFT in a black hole
  background}.
\newblock {\em JHEP}, 05:153, 2022.

\bibitem{Akal:2021dqt}
Ibrahim Akal, Taishi Kawamoto, Shan-Ming Ruan, Tadashi Takayanagi, and Zixia
  Wei.
\newblock {Page curve under final state projection}.
\newblock {\em Phys. Rev. D}, 105(12):126026, 2022.

\bibitem{Renner:2021qbe}
Renato Renner and Jinzhao Wang.
\newblock {The black hole information puzzle and the quantum de Finetti
  theorem}.
\newblock 10 2021.

\bibitem{Balasubramanian:2021xcm}
Vijay Balasubramanian, Ben Craps, Mikhail Khramtsov, and Edgar Shaghoulian.
\newblock {Submerging islands through thermalization}.
\newblock {\em JHEP}, 10:048, 2021.

\bibitem{Engelhardt:2022qts}
Netta Engelhardt and \r{A}smund Folkestad.
\newblock {Canonical purification of evaporating black holes}.
\newblock {\em Phys. Rev. D}, 105(8):086010, 2022.

\bibitem{Afrasiar:2022ebi}
Mir Afrasiar, Jaydeep Kumar~Basak, Ashish Chandra, and Gautam Sengupta.
\newblock {Islands for Entanglement Negativity in Communicating Black Holes}.
\newblock {\em arXiv:2205.07903}, 5 2022.

\bibitem{Fonda:2014cca}
Piermarco Fonda, Luca Giomi, Alberto Salvio, and Erik Tonni.
\newblock {On shape dependence of holographic mutual information in AdS$_{4}$}.
\newblock {\em JHEP}, 02:005, 2015.

\bibitem{Fonda:2015nma}
Piermarco Fonda, Domenico Seminara, and Erik Tonni.
\newblock {On shape dependence of holographic entanglement entropy in
  AdS$_{4}$/CFT$_{3}$}.
\newblock {\em JHEP}, 12:037, 2015.

\bibitem{Seminara:2017hhh}
Domenico Seminara, Jacopo Sisti, and Erik Tonni.
\newblock {Corner contributions to holographic entanglement entropy in
  AdS$_{4}$/BCFT$_{3}$}.
\newblock {\em JHEP}, 11:076, 2017.

\bibitem{Seminara:2018pmr}
Domenico Seminara, Jacopo Sisti, and Erik Tonni.
\newblock {Holographic entanglement entropy in AdS$_{4}$/BCFT$_{3}$ and the
  Willmore functional}.
\newblock {\em JHEP}, 08:164, 2018.

\bibitem{Cavini:2019wyb}
Giacomo Cavini, Domenico Seminara, Jacopo Sisti, and Erik Tonni.
\newblock {On shape dependence of holographic entanglement entropy in
  AdS$_{4}$/CFT$_{3}$ with Lifshitz scaling and hyperscaling violation}.
\newblock {\em JHEP}, 02:172, 2020.

\bibitem{brakke1992surface}
Kenneth~A Brakke.
\newblock The surface evolver.
\newblock {\em Experimental mathematics}, 1(2):141--165, 1992.

\bibitem{evolverlink}
Kenneth~A Brakke.
\newblock {\it Surface Evolver} program.
\newblock \url{http://www.susqu.edu/brakke/evolver/evolver.html}.

\bibitem{Randall:1999vf}
Lisa Randall and Raman Sundrum.
\newblock {An Alternative to compactification}.
\newblock {\em Phys. Rev. Lett.}, 83:4690--4693, 1999.

\bibitem{Dvali:2000hr}
G.R. Dvali, Gregory Gabadadze, and Massimo Porrati.
\newblock {4-D gravity on a brane in 5-D Minkowski space}.
\newblock {\em Phys. Lett. B}, 485:208--214, 2000.

\bibitem{Gubser:1999vj}
Steven~S. Gubser.
\newblock {AdS / CFT and gravity}.
\newblock {\em Phys. Rev. D}, 63:084017, 2001.

\bibitem{Miao:2017gyt}
Rong-Xin Miao, Chong-Sun Chu, and Wu-Zhong Guo.
\newblock {New proposal for a holographic boundary conformal field theory}.
\newblock {\em Phys. Rev. D}, 96(4):046005, 2017.

\bibitem{Chu:2017aab}
Chong-Sun Chu, Rong-Xin Miao, and Wu-Zhong Guo.
\newblock {On New Proposal for Holographic BCFT}.
\newblock {\em JHEP}, 04:089, 2017.

\bibitem{Ling:2021vxe}
Yi~Ling, Peng Liu, Yuxuan Liu, Chao Niu, Zhuo-Yu Xian, and Cheng-Yong Zhang.
\newblock {Reflected entropy in double holography}.
\newblock {\em JHEP}, 02:037, 2022.

\bibitem{Emparan:1999pm}
Roberto Emparan, Clifford~V. Johnson, and Robert~C. Myers.
\newblock {Surface terms as counterterms in the AdS / CFT correspondence}.
\newblock {\em Phys. Rev. D}, 60:104001, 1999.

\bibitem{Liu:2022pan}
Yuxuan Liu, Zhuo-Yu Xian, Cheng Peng, and Yi~Ling.
\newblock {Addendum to: Black holes entangled by radiation}.
\newblock {\em JHEP}, 11:043, 2022.

\bibitem{Izumi:2022opi}
Keisuke Izumi, Tetsuya Shiromizu, Kenta Suzuki, Tadashi Takayanagi, and
  Norihiro Tanahashi.
\newblock {Brane dynamics of holographic BCFTs}.
\newblock {\em JHEP}, 10:050, 2022.

\bibitem{Headrick:2009pv}
Matthew Headrick, Sam Kitchen, and Toby Wiseman.
\newblock {A New approach to static numerical relativity, and its application
  to Kaluza-Klein black holes}.
\newblock {\em Class. Quant. Grav.}, 27:035002, 2010.

\bibitem{Dias:2015nua}
Óscar~J.C. Dias, Jorge~E. Santos, and Benson Way.
\newblock {Numerical Methods for Finding Stationary Gravitational Solutions}.
\newblock {\em Class. Quant. Grav.}, 33(13):133001, 2016.

\bibitem{Janik:2006ft}
Romuald~A. Janik.
\newblock {Viscous plasma evolution from gravity using AdS/CFT}.
\newblock {\em Phys. Rev. Lett.}, 98:022302, 2007.

\bibitem{Chesler:2008hg}
Paul~M. Chesler and Laurence~G. Yaffe.
\newblock {Horizon formation and far-from-equilibrium isotropization in
  supersymmetric Yang-Mills plasma}.
\newblock {\em Phys. Rev. Lett.}, 102:211601, 2009.

\bibitem{Bhattacharyya:2009uu}
Sayantani Bhattacharyya and Shiraz Minwalla.
\newblock {Weak Field Black Hole Formation in Asymptotically AdS Spacetimes}.
\newblock {\em JHEP}, 09:034, 2009.

\bibitem{Bosch:2016vcp}
Pablo Bosch, Stephen~R. Green, and Luis Lehner.
\newblock {Nonlinear Evolution and Final Fate of Charged Anti\textendash{}de
  Sitter Black Hole Superradiant Instability}.
\newblock {\em Phys. Rev. Lett.}, 116(14):141102, 2016.

\bibitem{Bosch:2019anc}
Pablo Bosch, Stephen~R. Green, Luis Lehner, and Hugo Roussille.
\newblock {Excited hairy black holes: Dynamical construction and level
  transitions}.
\newblock {\em Phys. Rev. D}, 102(4):044014, 2020.

\bibitem{Zhang:2021etr}
Cheng-Yong Zhang, Peng Liu, Yunqi Liu, Chao Niu, and Bin Wang.
\newblock {Dynamical charged black hole spontaneous scalarization in
  anti\textendash{}de Sitter spacetimes}.
\newblock {\em Phys. Rev. D}, 104(8):084089, 2021.

\bibitem{Zhang:2022cmu}
Cheng-Yong Zhang, Qian Chen, Yunqi Liu, Wen-Kun Luo, Yu~Tian, and Bin Wang.
\newblock {Dynamical transitions in scalarization and descalarization through
  black hole accretion}.
\newblock {\em Phys. Rev. D}, 106(6):L061501, 2022.

\bibitem{Chen:2022vag}
Qian Chen, Zhuan Ning, Yu~Tian, Bin Wang, and Cheng-Yong Zhang.
\newblock {Descalarization by quenching charged hairy black hole in
  asymptotically AdS spacetime}.
\newblock {\em JHEP}, 01:062, 2023.

\bibitem{Chen:2023eru}
Qian Chen, Zhuan Ning, Yu~Tian, Bin Wang, and Cheng-Yong Zhang.
\newblock {Nonlinear dynamics of hot, cold and bald Einstein-Maxwell-scalar
  black holes in AdS spacetime}.
\newblock 7 2023.

\bibitem{Janik:2017ykj}
Romuald~A. Janik, Jakub Jankowski, and Hesam Soltanpanahi.
\newblock {Real-Time dynamics and phase separation in a holographic first order
  phase transition}.
\newblock {\em Phys. Rev. Lett.}, 119(26):261601, 2017.

\bibitem{Chen:2022cwi}
Qian Chen, Yuxuan Liu, Yu~Tian, Bin Wang, Cheng-Yong Zhang, and Hongbao Zhang.
\newblock {Critical dynamics in holographic first-order phase transition}.
\newblock {\em JHEP}, 01:056, 2023.

\bibitem{Chen:2022tfy}
Qian Chen, Yuxuan Liu, Yu~Tian, Xiaoning Wu, and Hongbao Zhang.
\newblock {Quench Dynamics in Holographic First-Order Phase Transition}.
\newblock 11 2022.

\end{thebibliography}

\end{document}